\documentclass[12pt,tightenlines,eqsecnum,floats,aps,amsmath,amssymb,nofootinbib,prd]{revtex4}

\usepackage{setspace}
\usepackage{amsmath,amssymb,amsfonts,amsthm,mathrsfs}
\usepackage{graphicx}
\usepackage{enumerate}
\usepackage{subfigure}
\usepackage{color}

\begin{document}

\title{Power spectrum of primordial perturbations for an emergent universe in quantum reduced loop gravity}

\author{Javier Olmedo$^{1,2}$ and Emanuele Alesci$^{1}$}
\affiliation {1. Institute for Gravitation and the Cosmos, Penn State University,
  University Park, PA 16801 \\2. Department of Physics and Astronomy, Louisiana State University, Baton Rouge, LA 70803-4001, USA
}

\begin{abstract}
We study the dynamics and predictions of a new emergent-universe model recently derived within Quantum Reduced Loop Gravity and based on the so-called statistical regularization scheme. These effective geometries show a dynamical transition from a stationary spacetime, with nearly constant scale factor at very early times, to a late-time semiclassical phase well approximated by a classical Friedmann-Robertson-Walker spacetime. We show that this is always the case when the matter content is a minimally coupled scalar field subject to a quadratic potential, including the massless case. Besides, a finite period of (nearly) exponential expansion in the semiclassical region can take place. Hence, we incorporate cosmological scalar and tensor perturbations, with a well-defined dynamics, and compute their power spectra at the end of inflation. We show that they are nearly scale invariant up to some scale where scale invariance is broken. Besides, they show qualitative differences with respect to the bouncing scenario of Loop Quantum Cosmology at scales where the scale invariance is broken. Nevertheless, the tensor-to-scalar ratio remains approximately constant even for modes well affected by the background evolution. 
\end{abstract}
\maketitle

\section{Introduction}

Inflation is nowadays the paradigm for the physics of the early universe that provides the simplest and most accurate description that is compatible with the large-scale observations of the temperature anisotropies of the cosmic microwave background (CMB). High-precision cosmological missions like Planck \cite{planck-com,planck-inf} provide measurements compatible with a nearly scale-invariant primordial power spectrum. This is precisely the situation when inflation is driven by a scalar field with a (suitably chosen) potential. Although the energy scales of inflation are typically small compared with the Planck regime, where quantum gravity is expected to be relevant, it is very difficult to ignore the pre-inflationary epoch of our Universe. Unfortunately, there is no consensus yet on a complete and satisfactory quantum theory of gravity. Among the possible candidates, Loop Quantum Gravity (LQG) \cite{lqg,lqg1} is a quantization program based on a nonperturbative, background independent, canonical quantization of general relativity in real connection variables. Together with LQG, Loop Quantum Cosmology (LQC) \cite{lqc,lqc1} has emerged in the last decades as a robust quantum theory of the cosmos. It adopts LQG quantization techniques for the quantization of symmetry reduced models of the classical theory. Here, quantum geometry corrections provide a natural modification of GR in the deep quantum regime, where the classical singularity is avoided and replaced by a quantum bounce. It is interesting to notice that alternatives to inflation, like bouncing cosmologies \cite{bounce}, can be naturally explained by quantum gravity effects \cite{ed-bounce}, rather than postulating the existence of some exotic matter content. However, the predictions of several models within LQC have been studied mainly in the context of inflation. In all these cases \cite{hyb,drsd}, one starts incorporating additional perturbative degrees of freedom on these geometries, as cosmological perturbations, and combines the loop representation for the background with a standard Fock quantization for these inhomogeneities. A mathematically consistent quantization with a concrete characterization of the physical sector was proposed in Refs. \cite{hyb,hyb1}. However, this quantization is formal (the full quantum dynamics has been only studied for the background \cite{hyb3}). When the backreaction of the inhomogeneities is negligible, it is possible to derive an effective Hamiltonian for the perturbations \cite{hyb1,hyb2,hyb3} that incorporates quantum fluctuations of the background state. Interestingly, the quantum corrected equations of motion of the perturbations are equivalent to the ones of perturbations on an effective dressed geometry, as it was initially suggested in Refs. \cite{drsd,drsd1}, following an alternative derivation (see Ref. \cite{hyb4} for a comparison of these two approaches). Therefore, once the dynamics for the state of the background is specified, the evolution of the perturbations can be solved as in standard cosmology. Within the context of inflation, several proposals \cite{drsd1,hyb-pred,hyb-pred2,ag,ag2} show good agreement with observations, where the power spectrum at the end of inflation shows a region where it is nearly scale-invariant, for wavenumbers larger than a given threshold scale $k_{LQC}$. For wavenumbers below this scale, different choices of initial vacuum state yield power spectra either enhanced or suppressed. Interestingly, novel criteria for the selection of vacuum states at the very early Universe pick out states producing suppression \cite{hyb-pred,hyb-pred2,ag,ag2}. Let us also notice that, even if the states of the background geometry are not sharply peaked, the predictions seem to be robust \cite{aag}. 

Besides, in the last years, a lot of effort has been focused on the derivation of cosmological models from the full theory. These new scenarios preserve some of the basic kinematical properties of the full theory: the discretization of some quantum geometrical operators and the graph structure. Besides, in the situations where the new quantum corrections can be neglected, they show good agreement with LQC. Quantum Reduced Loop Gravity (QRLG) \cite{rev-qrlg} was an initial proposal that carries out a quantum symmetry reduction (rather than a classical one) after introducing a gauge fixing (with respect to the Gauss and momentum constraints) in the kinematical Hilbert space. On this gauge-fixed sector of the theory, one identifies suitable coherent states in the kinematical Hilbert space of LQG and computes the expectation value of the quantum Hamiltonian constraint. This expectation value provides an effective Hamiltonian that, in comparison with LQC, captures additional quantum structures from the full theory. For instance, in Ref. \cite{qrlg-lqc}, it was shown that the leading order contributions to the effective dynamics agree with those of LQC provided that the number of links in the graph can change. These studies provided additional insight about how some physically motivated choices made in LQC can be naturally interpreted from the point of view of the full theory. More refined proposals (based on the QRLG) for the derivation of symmetry reduced models from the full quantum theory have been recently studied \cite{dap-lie}. They take into consideration the regularization scheme of the Lorentzian part of the Hamiltonian constraint originally suggested by Thiemann \cite{lqg}. It turns out that the resulting effective cosmological spacetimes are bouncing cosmologies connecting a flat FRW cosmology to the future of the bounce with a de Sitter spacetime to its past, with a cosmological constant that is of the Planck order. Actually, some predictions  of this model have been recently discussed in Ref. \cite{agullo}. 

Interestingly, coherent states constructed out of superpositions of graphs with different number of edges have been seriously considered within so-called statistical regularization \cite{abcl,abs}. These new states introduce novel quantum corrections in the effective Hamiltonian with interesting properties: a detailed analysis of the effective dynamics showed that the resulting FRW spacetime geometry is not just a bouncing cosmology but an emergent universe. This kind of emergent universe scenarios was already discussed in the literature in the context of the classical theory \cite{emer}. However, they are not stable \cite{emer1}. It is worth to mention that some predictions of the statistical regularization scheme proposed within QRLG have been recently studied in Ref. \cite{barrau}.

In this manuscript, we will focus on the study of the dynamics of the effective emergent universe scenarios proposed in Refs. \cite{abcl,abs} and the extraction of predictions. We will start with a detailed description of the dynamics of these geometries, considering as matter content a minimally coupled scalar field subject to a quadratic potential and different choices of the mass, including the massless case. We will describe the properties of the quantum corrections in the different stages of the evolution as a perfect fluid. It will be completely characterized by an effective quantum energy density and pressure. We will compare these geometries with the ones in standard LQC and with the classical theory. Concretely, we will show that these spacetimes agree with LQC at late times (and therefore with GR). However, at very early times, they disagree. The classical theory approaches the classical singularity (at a finite proper time). In LQC the scale factor bounces once and continues the backward evolution into the collapsing branch. In QRLG, the scale factor bounces and recollapses (presumably infinitely many times), gradually reaching a constant value far in the past. There, the energy density of the scalar field approaches a constant value of the order of the Planck energy density. This is always the case for the different choices of matter content considered here. However, the pressure is either equal to the energy density (if the scalar field is massless) or it oscillates around zero, reaching a maximum magnitude whose value is of the order of the Planck energy density (provided the mass is nonvanishing). Actually, in the massive case, we show that a nearly exponential expansion at late times can take place within the emergent universe model. Hence, it is possible to study some predictions of these geometries within the context of slow-roll inflation and compare them with observations.

For this purpose, we will adopt the strategy of the classical theory \cite{bran} to these effective geometries. Namely, we will assume that quantum fluctuations of these effective spacetimes can be codified in quantum perturbations of the effective geometry. Therefore, we will assume that the cosmological perturbation theory is also valid here. Then, we introduce scalar and tensor perturbations and we neglect their backreaction (we adopt a test field approximation). Note that these perturbations are derived from the full classical action. Hence, we do not expect that neither they nor their equations of motion will capture all the physical properties of the actual fluctuations of the effective geometry. However, assuming that the effective geometries and their inhomogeneous quantum fluctuations are well described by their continuous limit, we only need to specify how quantum geometry corrections are going to be incorporated in their equations of motion by means of the homogeneous sector of the theory. Here, we will make a judicious choice, based on the following conditions: {\it i}) the evolution of each scalar and tensor mode should be well defined everywhere (smooth ordinary differential equations), {\it ii}) within a Fock quantization they admit a unitary dynamics, {\it iii}) they should agree with the classical equations of motion at late times, {\it iv}) the difference between the equations of motion of (the Mukhanov-Sasaki) scalar and tensor perturbations is codified in a potential $\cal U$ that vanishes when $m=0$, {\it v}) at early times the equations of motion of tensor perturbations should converge (dynamically) to the ones of a massless scalar field on Minkowski, and {\it vi}) they should be as simple as possible. We provide a set of equations that satisfy the previous requirements. Besides, our numerical investigations suggest that it is possible to recover a nearly scale-invariant power spectrum compatible with observations. However, our model introduces a scale (as LQC does) where the scale invariance is broken. This scale is of the same order of the one of LQC, however, the way in which scale invariance is broken is different. Remarkably, the tensor-to-scalar ratio, in agreement with LQC, remains nearly scale invariant, even at scales where the scale invariance of the power spectra is broken. 

This manuscript is organized as follows. In Sec. \ref{sec:prelim} we discuss the classical background spacetime and its quantum corrected counterpart within LQC. Our emergent model is discussed in Sec. \ref{sec:effective}. Cosmological perturbations and the primordial power spectrum are studied in Sec. \ref{sec:perts}. We conclude and discuss the results in Sec. \ref{sec:discuss}. Besides, for the sake of completeness, we have added two appendixes.

\section{Preliminaries}\label{sec:prelim}

We will consider here homogeneous and isotropic geometries with compact and flat spatial topologies (isomorphic to a three-torus). Typically, the spacetime metric is given by 
\begin{equation}
ds^2=-N^2(t)dt^2+a^2(t)d\vec x^2,
\end{equation}
where $x_i\in [0,l_0]$ are suitable coordinates well adapted to the spacetime symmetries, with $l_0$ denoting the maximum coordinate length on each spatial direction, $N(t)$ is a homogeneous lapse function and $a(t)$ is the dimensionless scale factor that only depends on the time $t$. The matter content will be a scalar field $\phi$ subject to the quadratic potential $V(\phi) = \frac{1}{2}m^2\phi^2$. 

The Hamiltonian formalism is the most convenient approach in our study. The matter sector will be described by the scalar field $\phi$ and its conjugate momentum denoted by $\pi_\phi$, such that $\{\phi,\pi_\phi\}=1$. In the geometrical sector, on the other hand, the natural variables in LQG will be the $su(2)$ real connection and densitized triad. However, following recent studies in LQC, it is more convenient to choose as geometrical variables the physical volume $v=v_0a^3$, with $v_0=l_0^3$, and the conjugate variable $b$, such that $\{b,v\}=4\pi G \gamma$.

The classical dynamics is determined by the homogeneous constraint 
\begin{equation}\label{eq:class-ham}
H_{\rm cl}(N_{\rm cl}) =  N_{\rm cl}(H^{\rm class}_{\rm gr}+H_{\rm matt})
\end{equation}
where
\begin{equation}\label{eq:ham-class-gr}
H^{\rm cl}_{\rm gr}=-\frac{3}{16 \pi G \gamma^2} vb^2,
\end{equation}
and
\begin{equation}\label{eq:ham-matt}
H_{\rm matt}=\frac{\pi_\phi^2}{2v}+\frac{1}{2}vm^2\phi^2.
\end{equation} 
The equations of motion can be easily computed by means of the Hamilton equations $\dot q_i = \{q_i,H\}$ and $p_i = \{p_i,H\}$, for the configuration and momenta, respectively. In proper time, we combine the equations of motion of the physical volume and its conjugate momenta into the Friedmann and Raychaudhuri equations, namely,
\begin{equation}
H^2=\frac{8\pi G}{3}\rho,\quad \dot H=-4\pi G(P+\rho), 
\end{equation}  
where 
\begin{equation}\label{eq:matter}
\rho = \frac{\dot\phi^2}{2}+\frac{1}{2}m^2\phi^2,\quad P = \frac{\dot\phi^2}{2}-\frac{1}{2}m^2\phi^2,  
\end{equation}  
are the energy density and the pressure of the scalar field, respectively, and $H=\frac{\dot a}{a} = \frac{1}{3} \frac{\dot v}{v}$ is the Hubble parameter. In addition, we combine the equations of motion of the scalar field and its momentum into the second order ordinary differential equation
\begin{equation}\label{eq:2nd-diff-phi}
\ddot\phi+3 H\dot\phi+\frac{\partial V(\phi)}{\partial\phi}=0,
\end{equation}

Typically, the Friedmann equation is regarded as a constraint on the initial data, while the Raychaudhuri and the scalar field equations of motion as second-order ordinary differential equations that must be integrated in order to solve the dynamics. In GR, given the mass of the scalar field, one needs to specify the value of the scalar field and the Hubble rate as initial data. One can fix the value of the scale factor to be equal to 1 at any arbitrary time (typically today). We determine the magnitude of the velocity of the scalar field, namely $\dot\phi$, via the Friedmann equation, while we fix its sign. 

In LQC \cite{lqc,lqc1}, one starts with a quantum version of the classical kinematical algebra $\{e^{i \lambda b},v\}=i 4\lambda\pi G \gamma e^{i \lambda b}$, for some constant $\lambda$, represented in the quantum kinematical Hilbert space by means of an appropriate operator representation. The basic variables will be holonomies of the connection and fluxes of the triad. On the kinematical Hilbert space, the eigenstates of the triad operator $\{|p\rangle ,p\in \mathbb{R}\}$ provide a basis of normalized states with respect to the discrete inner product $\langle p|p'\rangle=\delta_{p,p'}$. 

The Hamiltonian constraint operator determines the quantum dynamics. The physical states can be computed (exactly for $V(\phi)=0$), along with suitable Dirac observables and a physical inner product. Then, we can compute the expectation values of physical operators on sharply peaked states and track their trajectories. It is then natural to assume that, when the potential energy of the scalar field can be neglected against the its kinetic energy at the bounce, the LQC effective dynamics will reproduce accurately the actual trajectories of the expectation values of suitable operators. The effective Friedmann and Raychaudhuri equations take the form   
\begin{equation}
H^2=\frac{8\pi G}{3}\rho\left(1-\frac{\rho}{\rho_c}\right),
\end{equation}
and
\begin{equation}
\dot H =-4\pi G (P+\rho) \left(1-2\frac{\rho}{\rho_c}\right),
\end{equation}  
respectively. Here, $\rho$ and $P$ are still given by Eq. \eqref{eq:matter}, $\rho_c=\frac{3}{8\pi G\gamma^2 \lambda^2}$, $\lambda^2 = 4\sqrt{3}\pi\gamma\ell_{\rm Pl}^2$ and $\gamma$ is the Immirzi parameter of LQG. The equations of motion of the scalar field remain the same as in the classical theory -- see Eq. \eqref{eq:2nd-diff-phi}. 

The quantum corrections in those equations, although they come from the geometrical sector, have been expressed as an effective matter content. The effective quantum energy density in LQC is defined as
\begin{equation}\label{eq:q-rho-lqc}
\rho_{Q_{LQC}}=-\frac{\rho^2}{\rho_c}.
\end{equation}
In addition, we can also define the effective quantum pressure in LQC as
\begin{equation}\label{eq:q-press-lqc}
P_{Q_{LQC}}=-(\rho + 2 P)\frac{\rho}{\rho_c}.
\end{equation}  

As in GR, the Friedmann equation is a constraint that constrains the initial data. However, these cosmological models have a privileged space-like surface: the cosmological bounce. Here, the Hubble parameter vanishes. Therefore, one only needs to specify as initial data the value of the scalar field, namely $\phi_B$. At this bounce surface, we fix the scale factor to be equal to 1 and we determine the magnitude of $\dot\phi$ via the Friedmann equation. As in GR, we must provide its sign. Therefore, in LQC, the initial value problem is simpler than in GR. One only needs to specify $\phi_B$ at the bounce and if it is rolling up or down the potential there.

\section{Effective Hamiltonian, equations of motion and their solutions}\label{sec:effective}

Interestingly, a recently suggested dynamical scheme within QRLG, denoted as the statistical regularization scheme, shows very interesting properties. On the one hand, the typical dynamical schemes that mimic the dynamics of isotropic LQC models can be obtained in QRLG from kinematical coherent states where the underlying graphs have either a fixed number of edges $N^3$ ($N$ if we consider only one spatial direction) or a fixed quantum number $j$ (label of the $SU(2)$ irreducible representations whose nonvanishing minimum is $j = 1/2$). On the other hand, the new statistical regularization scheme incorporates contributions of superposition of spin networks with different $N$ and $j$ that collectively represent the same geometry.

Within this new scheme, one considers as kinematical states suitable density matrices superposing coherent spin networks representing the same cosmological model but with a different number of edges. Concretely, one assumes a superposition that follows a binomial distribution (indistinguishable graphs for a given $N^3$) from $N_{\rm min}^3=1$ to some $N^3_{\rm max}$ that grows with the physical volume $v$. For states with large $N_{\rm max}^3$, the binomial distribution can be replaced by a Gaussian distribution peaked around large values of $N^3$, which also means that one can take the continuum limit and replace the discrete sums by integrals. For details, see Ref. \cite{abs}.

In total, the effective Hamiltonian of the gravitational sector is obtained as the expectation value of the gravitational Hamiltonian constraint of LQG on this state. It takes the form
\begin{equation}\label{eq:hgr-qrlg}
H^{\rm gr} = -\frac{3}{8\pi G\gamma^2} v^{1/3}\int_{1}^{2v/\lambda^3}\frac{dN}{\sqrt{\pi v/\lambda^3}}e^{-\frac{\left(N-v/\lambda^3\right)^2}{v/\lambda^3}}N^{2/3}\sin^2\left(\frac{b\,v^{1/3}}{N^{1/3}}\right).
\end{equation}
For the matter sector, we assume the same Hamiltonian given in Eq. \eqref{eq:ham-matt}.

The equation of motion for $v$ is
\begin{equation}
\dot{v}=-4\pi G\gamma \frac{\partial H^{\rm tot}}{\partial b} = \frac{3v}{2\sqrt{\pi}\gamma\lambda}I_1(v,b),
\end{equation}
where the dot indicates derivative with respect to proper time, and where we have defined
\begin{equation}\label{eq:I1}
I_1(v,b)=\int_{x_{min}}^{x_{max}}dx \,e^{-x^2} \left(1+\frac{x}{\sqrt{\sigma}}\right)^{1/3}\sin\left[\frac{2\lambda b}{\left(1+\frac{x}{\sqrt{\sigma}}\right)^{1/3}}\right],
\end{equation}
with
\begin{equation}
x=\frac{\left(N-\sigma\right)}{\sqrt{\sigma}}, \quad \sigma=\frac{v}{\lambda^3},
\end{equation}
and 
\begin{equation}
x_{min} = -\sqrt{\sigma}+\frac{1}{\sqrt{\sigma}},\quad x_{max} = \sqrt{\sigma}.
\end{equation} 

The equation of motion for $b$ is given by
\begin{align}\nonumber
\dot b &= 4\pi G\gamma \frac{\partial H^{\rm tot}}{\partial v} = \frac{\pi G \gamma }{3}\left(14 V(\phi)-\frac{5p_\phi^2}{v^2}\right)
- 3\frac{2^{2/3}}{\sqrt{\pi}\gamma \lambda^2 } \sqrt{\sigma}e^{-\sigma} \sin^2\left[\frac{\lambda b}{2^{1/3}}\right]\\ \label{eq:dotb}
&- \frac{b}{2\sqrt{\pi}\gamma\lambda} I_1(v,b) - \frac{3\sigma}{2\sqrt{\pi}\gamma\lambda^2} I_2(v,b).
\end{align}
with the phase space function
\begin{equation}\label{eq:I2}
I_2(v,b) = \int_{x_{min}}^{x_{max}}dx \,e^{-x^2} \left(1+\frac{x}{\sqrt{\sigma}}\right)^{2/3}\sin^2\left[\frac{\lambda b}{\left(1+\frac{x}{\sqrt{\sigma}}\right)^{1/3}}\right]\left(\frac{x^2}{\sigma}+\frac{2 x}{\sqrt{\sigma}}\right).
\end{equation}
Let us notice that in Eq. \eqref{eq:dotb} we have used the Hamiltonian constraint once in order to simplify some terms. Again, the equations of motion of the scalar field can be combined into the second order differential equation \eqref{eq:2nd-diff-phi}, taking exactly the same explicit form as in the classical theory.

In this manuscript, we will discuss in some detail two cases: i) $m=0$, i.e. massless scalar field, and ii) $m\neq 0$. The case $m=0$ was considered in Refs. \cite{abcl,abs} under the saddle-point approximation for the integral in Eq. \eqref{eq:hgr-qrlg}. However, the analysis was limited and it emphasized the emergent nature of the spacetime without studying its dynamics in detail. The case $m\neq 0$ has been partially analyzed in Ref. \cite{barrau} for values of the mass motivated by the constraints in observations. However, several interesting aspects of the dynamics were not discussed there, likely because of that restriction in the values of the mass together with the limitations of the numerical analysis. 

Here, we will carry out a more sophisticated numerical study of the dynamics, without restricting the values of the mass phenomenologically. The main assumption will be the approximation of the series (summation in the number of vertices $N^3$) by continuous integrals at early times, without introducing any saddle-point approximations. Only when the trajectories are well inside the classical region we continue the evolution with the simplified equations given in Ref. \cite{abcl,abs}. The reason for this is because we lose precision in the numerical evaluation of the integrals at very large volumes \footnote{Actually, some trajectories can reach very large volumes even in the quantum region. Here, we do not explore these trajectories in detail since our numerical tools, at the moment, are not appropriate for them.}. 

Below we provide our numerical analysis. For our simulations, $\gamma=0{.}2375$, $G=1=c$ and $\hbar=1$ (Planck units). In the following, unless otherwise is specified, all numerical values will be given in Planck units. Besides, as we already mentioned, due to the challenge that represents to reach the required precision in the numerical integration of the previous set of equations at large volumes (concretely an accurate estimation of the integrals), at $b= 0{.}69088\cdot 10^{-2}$, well in the classical regime, we continue the forward evolution within the saddle-point approximation for the integrals, as it was already done in Refs. \cite{abcl,abs}. The concrete Hamiltonian and set of equations of motion are provided in App. \ref{app:sp-appr}. We have checked that the matching is smooth and, therefore, the error under control. We have also checked that we obtain similar results if instead the matching is done with LQC equations of motion. Moreover, in order to compare the solutions of our approach (QRLG) with LQC and with the classical theory, we match them well in the classical region, where the three approaches agree very well. Those readers interested in more concrete details about our numerical simulations should see App. \ref{app:numer}.

\subsection{The massless scalar field model}

\begin{figure}[h]
{\centering     
\includegraphics[width = 0.8\textwidth]{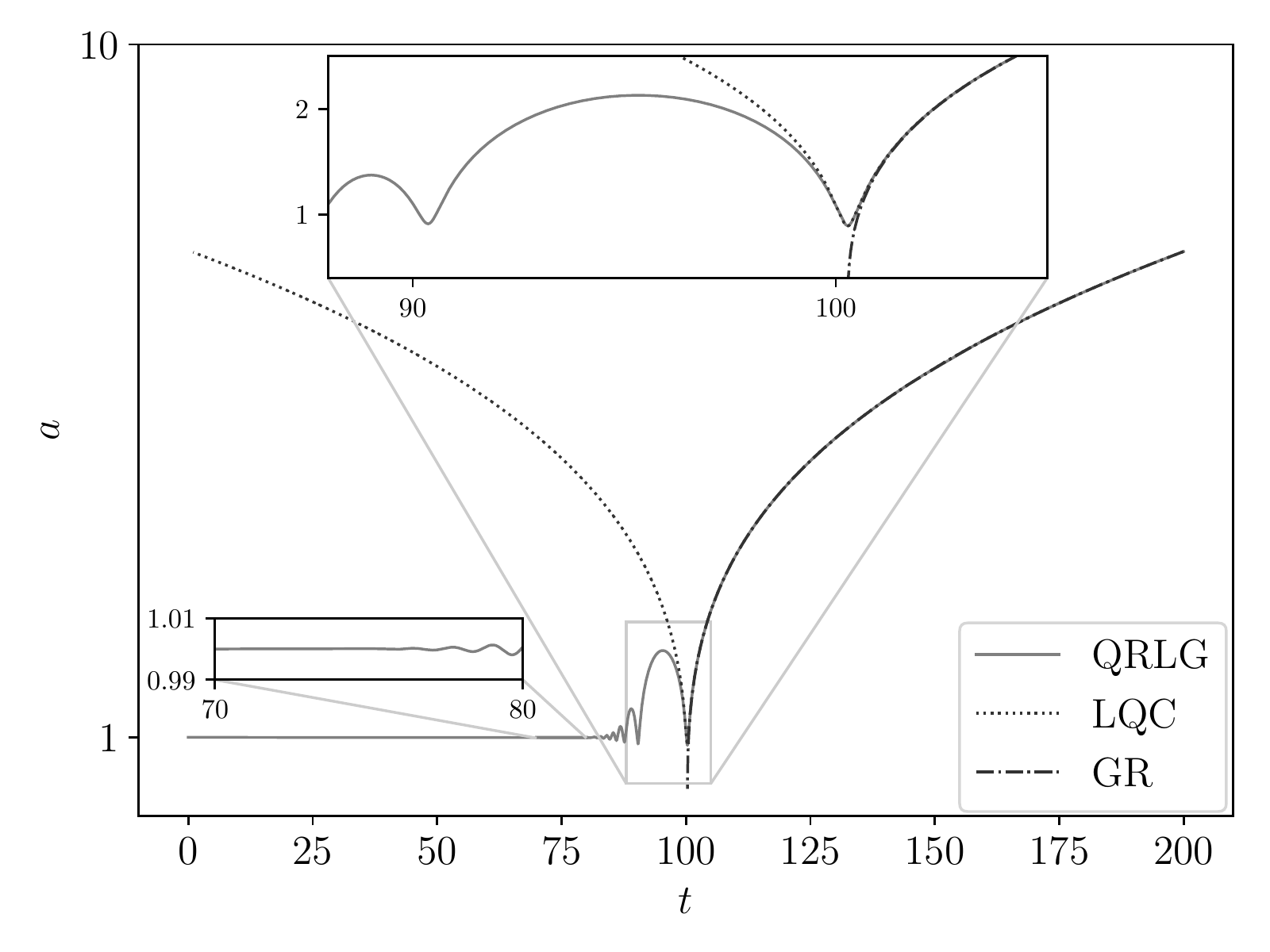}  
}
\caption{This graph shows the scale factor computed within the effective dynamics of our model, the effective geometries obtained in LQC and the classical theory. In all cases, we match the trajectories at late times (where the classical theory is a good approximation). The initial data at $t_0=0$ is $v(t_0)=125.0$, $b(t_0)=1.09\cdot10^2$, $\phi(t_0)=0$ and $\dot\phi(t_0)>0$ with its value determined by solving the Hamiltonian constraint.}
\label{fig:a-m0}
\end{figure}

Most of the properties of these effective geometries can be discussed for the simple case $V(\phi)=0$. In Fig. \ref{fig:a-m0} we show the scale factor for our effective model within QRLG, the effective dynamics of LQC and the classical trajectory in GR. We find good agreement in the three cases few Planck seconds after the last bounce. However, when quantum corrections become important, the three descriptions disagree. In particular, the scale factor in the classical theory goes to zero quickly (hitting the classical singularity). The trajectory described by our model and the one in LQC share some qualitative properties around the first bounce, which in both cases occurs approximately at the same time. For smaller times, we see that the LQC trajectory continues its way (backwards) to the collapsing branch, while in our model the scale factor remains in a strong quantum regime, bouncing and recollapsing, and reaching a constant (Planck order) magnitude in the asymptotic past.

\begin{figure}
{\centering     
\includegraphics[width = 0.8\textwidth]{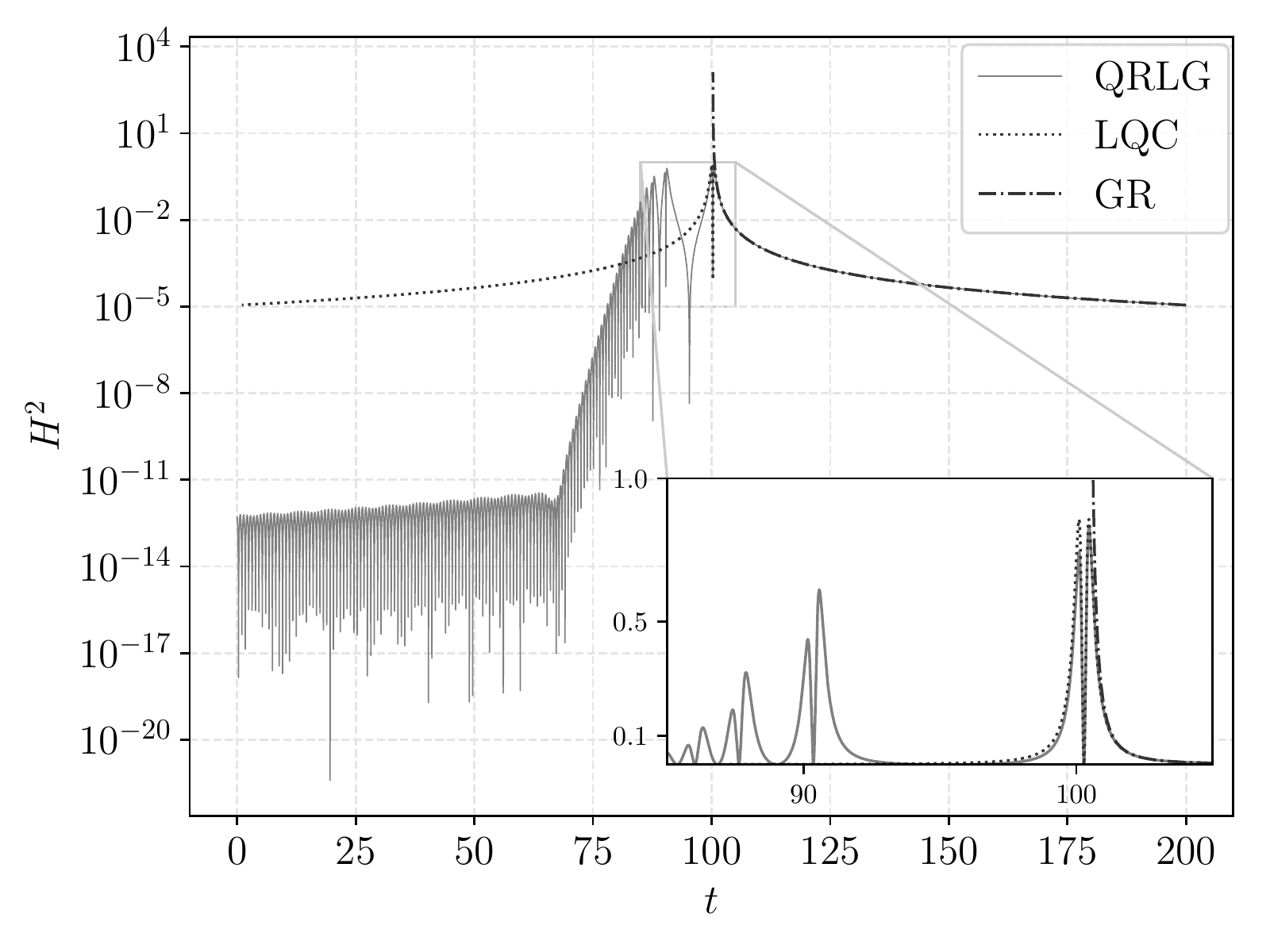}\\  
\includegraphics[width = 0.8\textwidth]{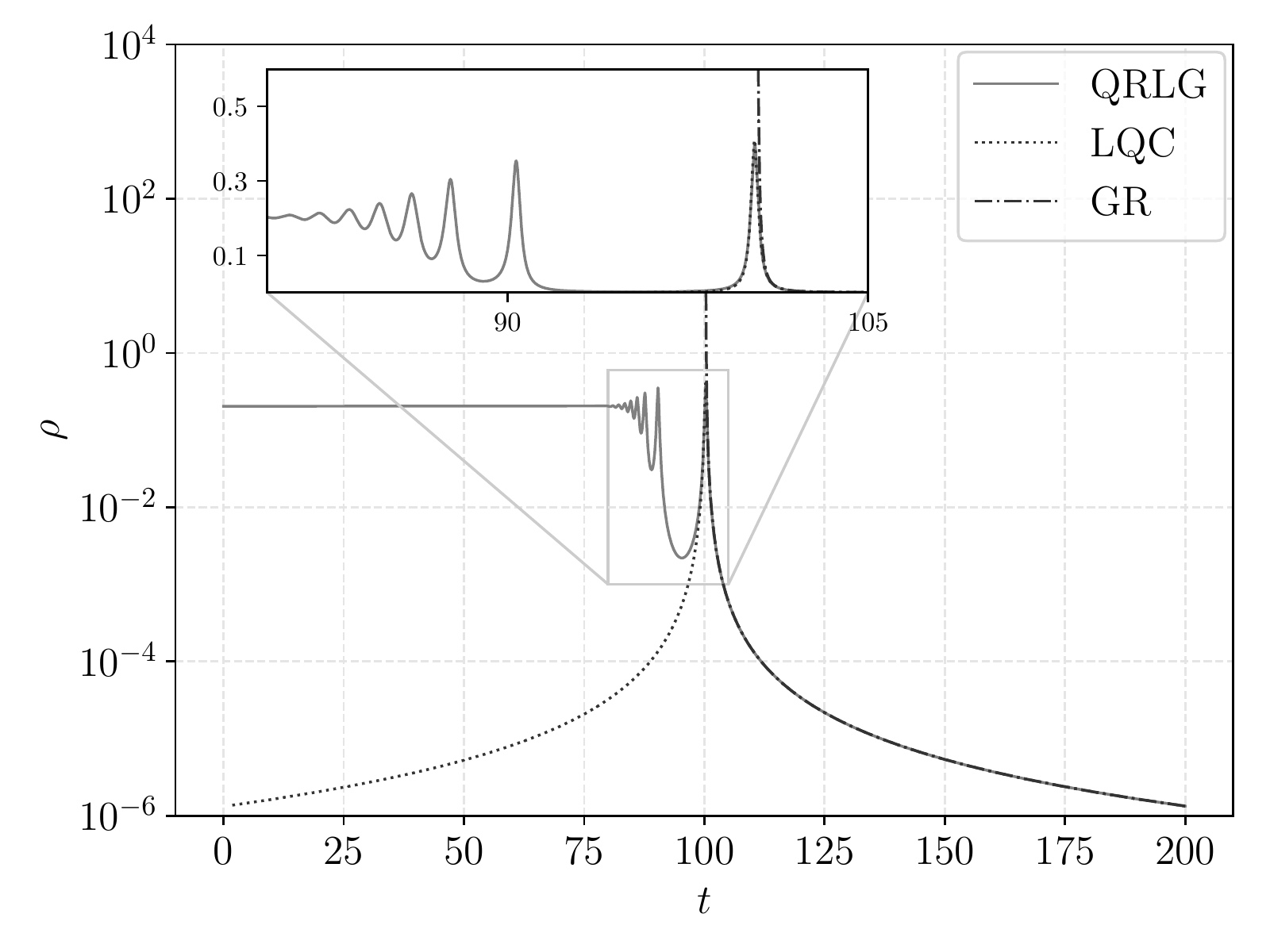}}
\caption{These graphs provide a comparison of the (square of the) Hubble parameter (upper panel) and the energy density of the scalar fields (lower panel) computed within the effective dynamics of our model, the effective geometries obtained in loop quantum cosmology and the classical theory. Here, we match the trajectories at late times (where the classical theory is a good approximation). The initial data at $t_0=0$ is $v(t_0)=125.0$, $b(t_0)=1.09\cdot10^2$, $\phi(t_0)=0$ and $\dot\phi(t_0)>0$ with its value determined by solving the Hamiltonian constraint.}
\label{fig:H-rho-m0}
\end{figure}

A closer look at the Hubble parameter, as shown in the upper panel of Fig. \ref{fig:H-rho-m0}, confirms these statements. Again, away from the quantum regions, we find good agreement between the classical theory, LQC and our model. When the quantum corrections become important, the classical Hubble parameter diverges, while in the other approaches it remains finite. As we saw in Fig. \ref{fig:a-m0} for the scale factor, at the first bounce, LQC and our model show good agreement regarding the instants at which the square of the Hubble parameter reaches its maximum and minimum values. However, they disagree in magnitude. Then, in the LQC trajectory, the Hubble parameter decreases back in time in the collapsing branch (in good agreement with GR). Interestingly, in our model, the Hubble parameter oscillates very fast around zero (although this is not shown in Fig. \ref{fig:H-rho-m0} since we plot $H^2$), with an amplitude that decreases monotonously. 

In the lower panel of Fig. \ref{fig:H-rho-m0} we see that the energy density of our model agrees with the energy density of the classical trajectory at late times (so does the energy density for the effective LQC). However, as expected, close to the first bounce, while the classical energy density diverges, the energy density of our model remains bounded above, as in LQC. Nevertheless, these upper bounds in LQC and our model disagree. Besides, while the LQC energy density decreases back in time at early times, in our model it remains approximately constant (except for small oscillations that are damped at early times). This behavior is in agreement with our previous discussion for the scale factor and the Hubble parameter, although it is not obvious. We will explain why this is the case in the next subsection. 

\begin{figure}
{\centering     
\includegraphics[width = 0.8\textwidth]{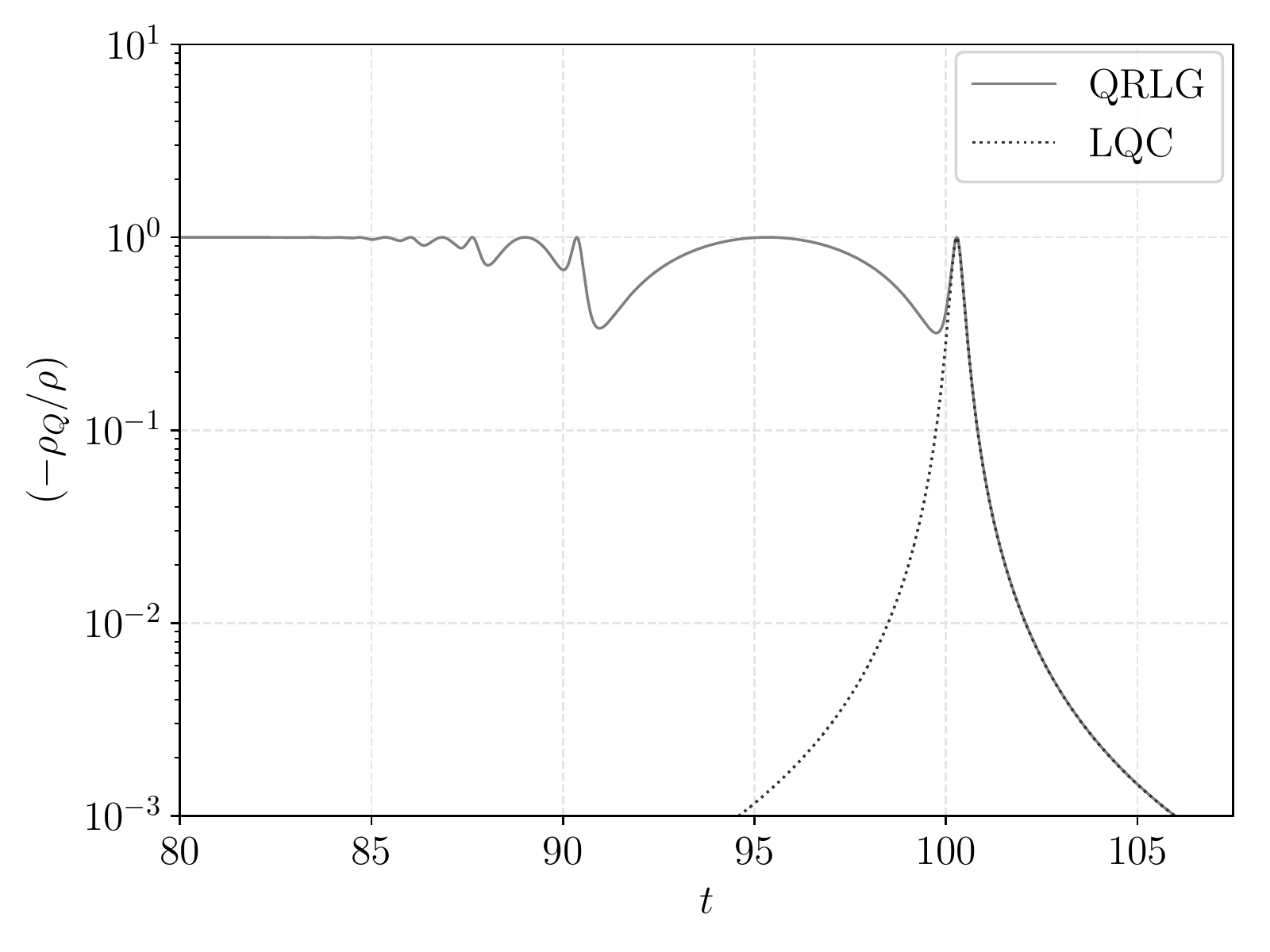}  
\includegraphics[width = 0.8\textwidth]{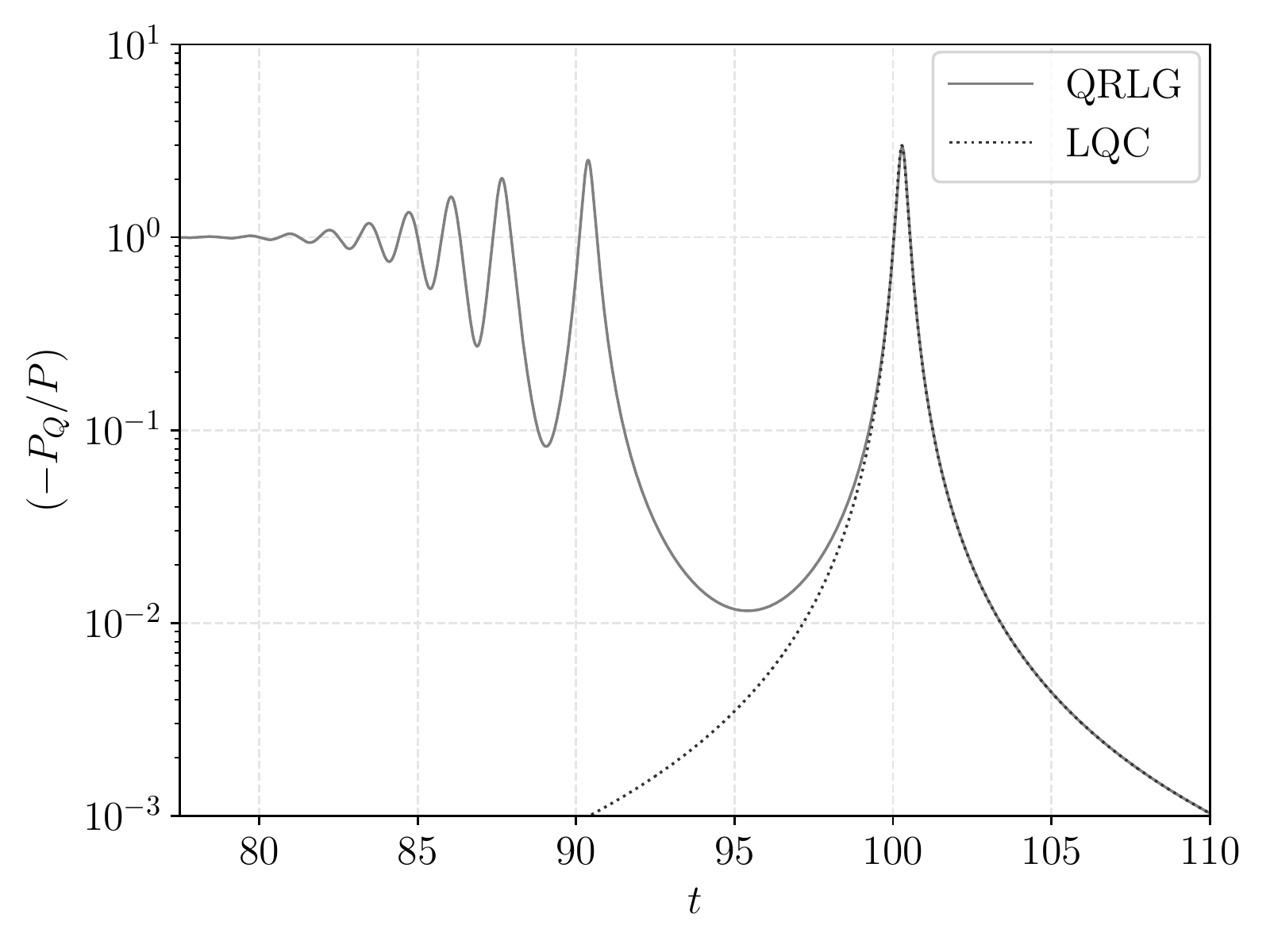}}
\caption{In these graphs, we compare the effective trajectories of our model and those of loop quantum cosmology, the ratio between the effective quantum energy density and the energy density of the scalar field (upper panel) and the ratio between the effective quantum pressure and the pressure of the scalar field (lower panel). We match the trajectories at late times (where general relativity is a good approximation). The initial data at $t_0=0$ is $v(t_0)=125.0$, $b(t_0)=1.09\cdot10^2$, $\phi(t_0)=0$ and $\dot\phi(t_0)>0$ with its value determined by solving the Hamiltonian constraint.}
\label{fig:rho-P-m0}
\end{figure}

Now, as is customary in quantum cosmology, we treat the quantum corrections that modify Einstein equations as an effective fluid in classical general relativity. Let us start with the Friedmann equation. We define the effective quantum energy density as 
\begin{equation}\label{eq:q-rho}
\rho_Q=\frac{3}{8\pi G} H^2-\rho.
\end{equation}  
Let us recall that, in LQC, the effective quantum energy density is given in Eq. \eqref{eq:q-rho-lqc}. It is clear that, in those regimes where the classical dynamics holds, the ratio $\rho_Q/\rho_{\rm Pl}$ must be negligible. The left panel of Fig. \ref{fig:rho-P-m0} confirms this statement for LQC and our model. There, the effective quantum energy densities in both cases overlap except for numerical errors. At early times, on the other hand, $\rho_{Q_{LQC}}$ decreases fast back in time, while $\rho_{Q}$ remains constant, negative, and with a magnitude similar to the one of $\rho$. Therefore, in that regime, we have $\rho_Q+\rho\simeq 0$. The consequence of this surprising behavior is that $H^2\simeq 0$.

Let us focus now on the Raychaudhuri equation. We define the effective quantum pressure as
\begin{equation}
P_Q=-\frac{1}{4\pi G} \dot H-(P+\rho+\rho_Q),
\end{equation}  
with the time derivative of the Hubble parameter with respect to proper time given by
\begin{equation}
\dot H=- \frac{5}{2}H^2 + \frac{3}{2 \sqrt{\pi} \gamma \lambda}  \left(2^{4/3} H \sqrt{\sigma} e^{-\sigma} \sin(2^{2/3} \lambda b) + I_3(v,b)\,H +\frac{2\lambda}{3} I_4(v,b)\, (\dot b + H b)\right),
\end{equation}
where we have introduced the phase space functions
\begin{equation}\label{eq:I3}
I_3(v,b) = \int_{x_{min}}^{x_{max}}dx \,e^{-x^2} \left(1+\frac{x}{\sqrt{\sigma}}\right)^{1/3}\sin\left[\frac{2\lambda b}{\left(1+\frac{x}{\sqrt{\sigma}}\right)^{1/3}}\right]\left(x^2+2 x\sqrt{\sigma}\right),
\end{equation}
and 
\begin{equation}\label{eq:I4}
I_4(v,b) = \int_{x_{min}}^{x_{max}}dx \,e^{-x^2} \cos\left[\frac{2\lambda b}{\left(1+\frac{x}{\sqrt{\sigma}}\right)^{1/3}}\right],
\end{equation}
with $\dot b$ given in Eq. \eqref{eq:dotb}. In LQC, Eq. \eqref{eq:q-press-lqc} gives the effective quantum pressure. We plot $P_Q$ in the lower panel of Fig. \ref{fig:rho-P-m0}. As we can see, at early times, the effective quantum pressure, up to oscillations whose amplitude decrease back in time, satisfies $P_Q + P\simeq0$. On the other hand, in LQC it reaches a maximum at the bounce and then decreases again as the trajectory enters the collapsing branch. Therefore, we see that at early times the effective quantum pressure of our model and the one of LQC disagree completely. Now, the behavior of the quantum pressure $P_Q$ together with the one of the quantum energy density $\rho_Q$ at early times explains why the scale factor takes a constant value: the effective quantum stress-energy tensor always counteract the matter energy density. Therefore, both the Hubble parameter and its time derivative will become negligible at very early times.

As a final remark, it is interesting to notice that, at late times, the effective quantum energy density and pressure of our model and the one of LQC are in very good agreement. In other words, at late times, the Friedmann and Raychaudhuri equations of our model and LQC agree very well. 

\subsection{Scalar field with a quadratic potential: dynamics and phenomenological aspects for observations}

Now, let us consider a nonvanishing mass for the scalar field. Before we move into our numerical analysis, we would like to discuss the qualitative behavior of the scalar field in the deep quantum region. Although this will be confirmed below for our model, let us consider the equation of motion of the scalar field given in Eq. \eqref{eq:2nd-diff-phi}. Let us assume that $\displaystyle \lim_{t\to-\infty} H= 0$, such that the friction term can be neglected. In addition, let us assume that the energy density reaches a constant value there, namely $\displaystyle \lim_{t\to-\infty}\rho(t) =\rho_0$. Then, one can easily conclude that at early times
\begin{equation}
\phi(t)\simeq\frac{\sqrt{2\rho_{0}}}{m}\sin(m\,t+\varphi_0),
\end{equation}
where $\varphi_0$ is some constant phase. In addition, we should notice that the pressure, given in Eq. \eqref{eq:matter}, will not be constant. One can easily see that
\begin{equation}\label{eq:press-t0}
P(t)\simeq\rho_{0}\cos(2 m\,t+\varphi_0).
\end{equation}

\begin{figure}
{\centering     
\includegraphics[width = 0.8\textwidth]{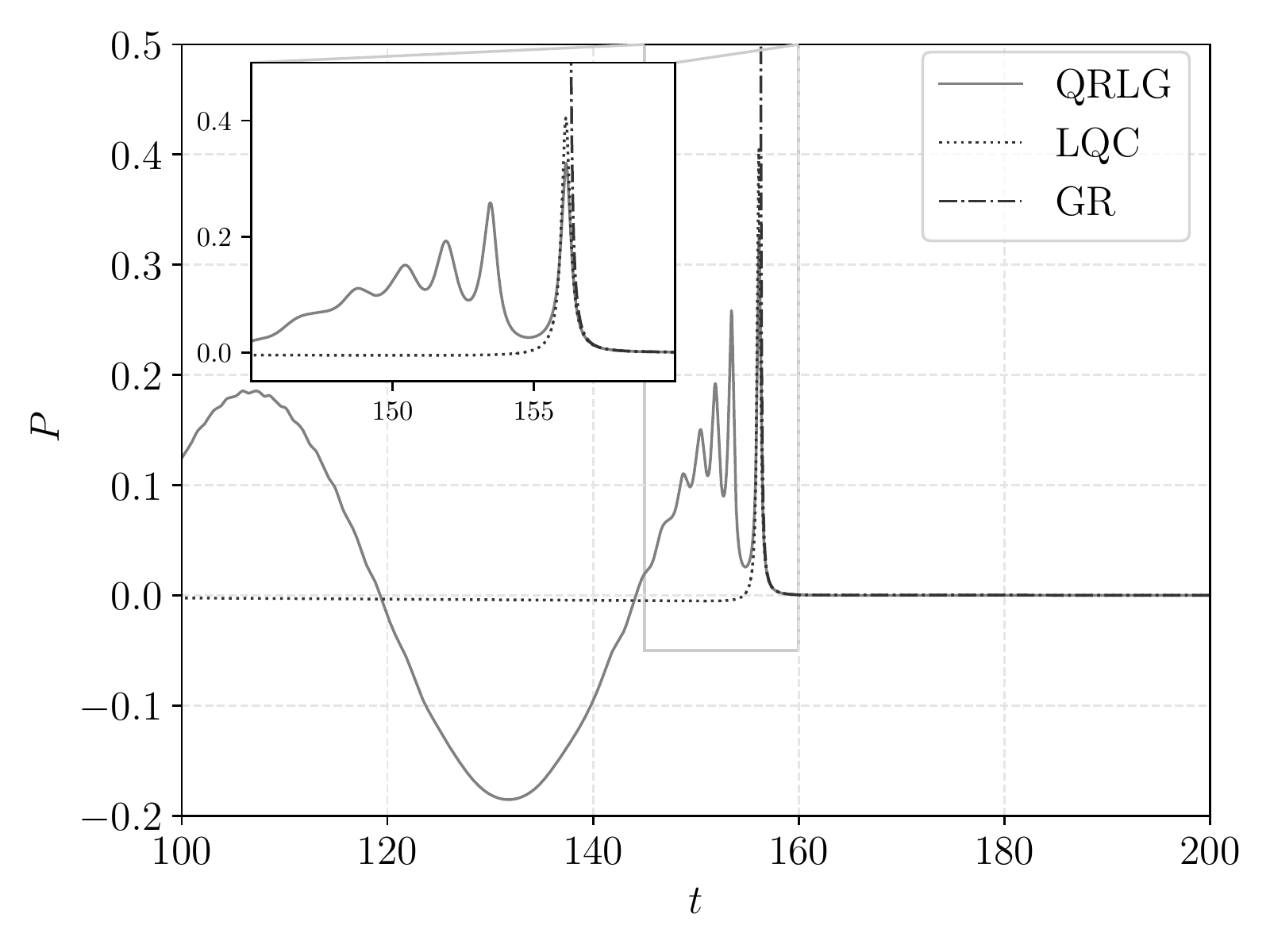}  
\includegraphics[width = 0.8\textwidth]{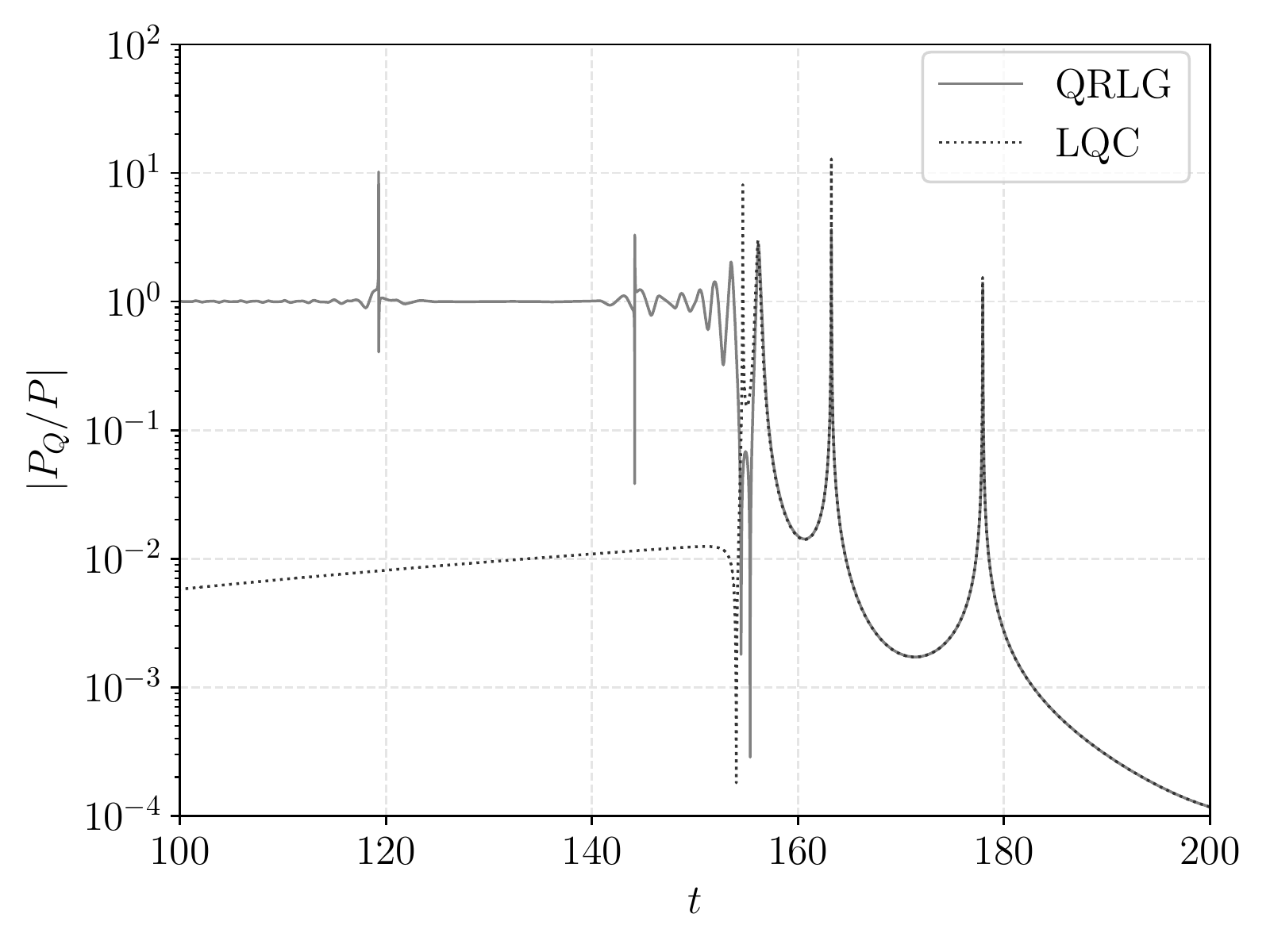}}
\caption{These graphs show a comparison of the pressure of the scalar field for the effective dynamics of our model, loop quantum cosmology and the classical theory (upper panel). We also compare the effective quantum pressure of our model and LQC (lower panel). Here, we match the trajectories at late times (where the classical theory is a good approximation). The mass in the potential is $m=6{.}25\cdot 10^{-2}$. The initial data at $t_0=0$ is $v(t_0)=27.0$, $b(t_0)=1.09\cdot10^2$, $\phi(t_0)=-3.7$ and $\dot\phi(t_0)>0$ with its value determined by solving the Hamiltonian constraint.}
\label{fig:P-m}
\end{figure}

Let us now discuss our numerical studies in more detail. Once again, they confirm that the emergent-universe scenario is robust even in the presence of a quadratic potential. We will consider a value of the mass that is not justified phenomenologically, but that allows us to understand several aspects that, to our knowledge, were not discussed in the literature before. On the one hand, we have seen that in this case, we get a similar behavior as in the case of a massless field: the energy density at asymptotic early times reaches a constant value while the Hubble parameter and its time derivative become very small there. We do not show the plot here since they qualitatively agree with the behavior already shown in Fig. \ref{fig:H-rho-m0} in the previous subsection. On the other hand, in the upper panel of Fig. \ref{fig:P-m} we compare the pressure of the scalar field $P$, for our model, LQC and the classical theory. We show this comparison because the behavior of $P$ when the mass is not zero is qualitatively different from the previous massless case. Again, at late times, we see good agreement between the dynamics of the three models (the quantum effective pressures in the lower panel of Fig. \ref{fig:P-m} becomes negligible). When we reach the quantum region, while the pressure in GR blows up, in both our model and LQC it reaches a finite maximum. Then, if we continue the evolution to the past, the pressure in LQC and our model disagree. In LQC it goes to zero, as the trajectory approaches the collapsing branch in the backward evolution, while in our model it oscillates around zero, reaching a maximum (Planck order) magnitude. In the lower panel of Fig. \ref{fig:P-m}, at these early times, the quantum pressure of LQC becomes negligible again, while in our model it fulfills $P+P_Q\simeq 0$. This behavior is in agreement with the massless case. Actually, we have checked that both of them are well approximated by Eq. \eqref{eq:press-t0}.  
Note that in the classical region, since $\rho_Q$ and $P_Q$ in our model and LQC agree very well (and are much smaller than $\rho$ and $P$, respectively), we conclude that the effective Friedmann and Raychaudhuri equations are in good agreement again (and consequently with classical GR). This again confirms that the effective quantum stress-energy tensor always counteracts the matter content in the remote past, forcing the scale factor to reach a stationary configuration.

Understanding and characterizing this behavior at very early times is essential if one wants to define a probability distribution in order to determine how likely is this model in order to produce a sufficiently large number of $e$-folds at the end of inflation \cite{barrau}.

Finally, let us consider the phenomenological aspects of the model. Since we are interested in the comparison of some of the predictions of this model with observations, we will fix the mass of the scalar field to be equal to $m=1{.}27795\cdot 10^{-6}$. We extract its value from the observed amplitude and spectral index of the primordial power spectrum by the Planck collaboration \cite{planck-inf}, namely,
\begin{equation}\label{eq:planck}
\log(10^{10}A_s(k_\star)) = 3{.}094\pm 0{.}034,\quad n_s(k_\star) = 0{.}9645\pm 0{.}0049, \quad k_\star = 0{.}05\,(\rm Mpc)^{-1}
\end{equation}
and the well-known formulae to leading order in the slow-roll approximation 
\begin{equation}
A_s(k_\star)=\hbar \frac{G H^2(t_\star)}{\pi \epsilon_V(t_\star)},\quad n_s(k_\star) = 1-4 \epsilon_V(t_\star).
\end{equation}
Here, $t_\star$ is the time at which the mode $k_\star$ leaves the horizon, namely $k_\star=a(t_\star)H(t_\star)$ and 
\begin{equation}
\epsilon_V=\frac{1}{16\pi G} \left(\frac{V'(\phi)}{V(\phi)}\right).
\end{equation}

\begin{figure}
{\centering     
\includegraphics[width = 0.8\textwidth]{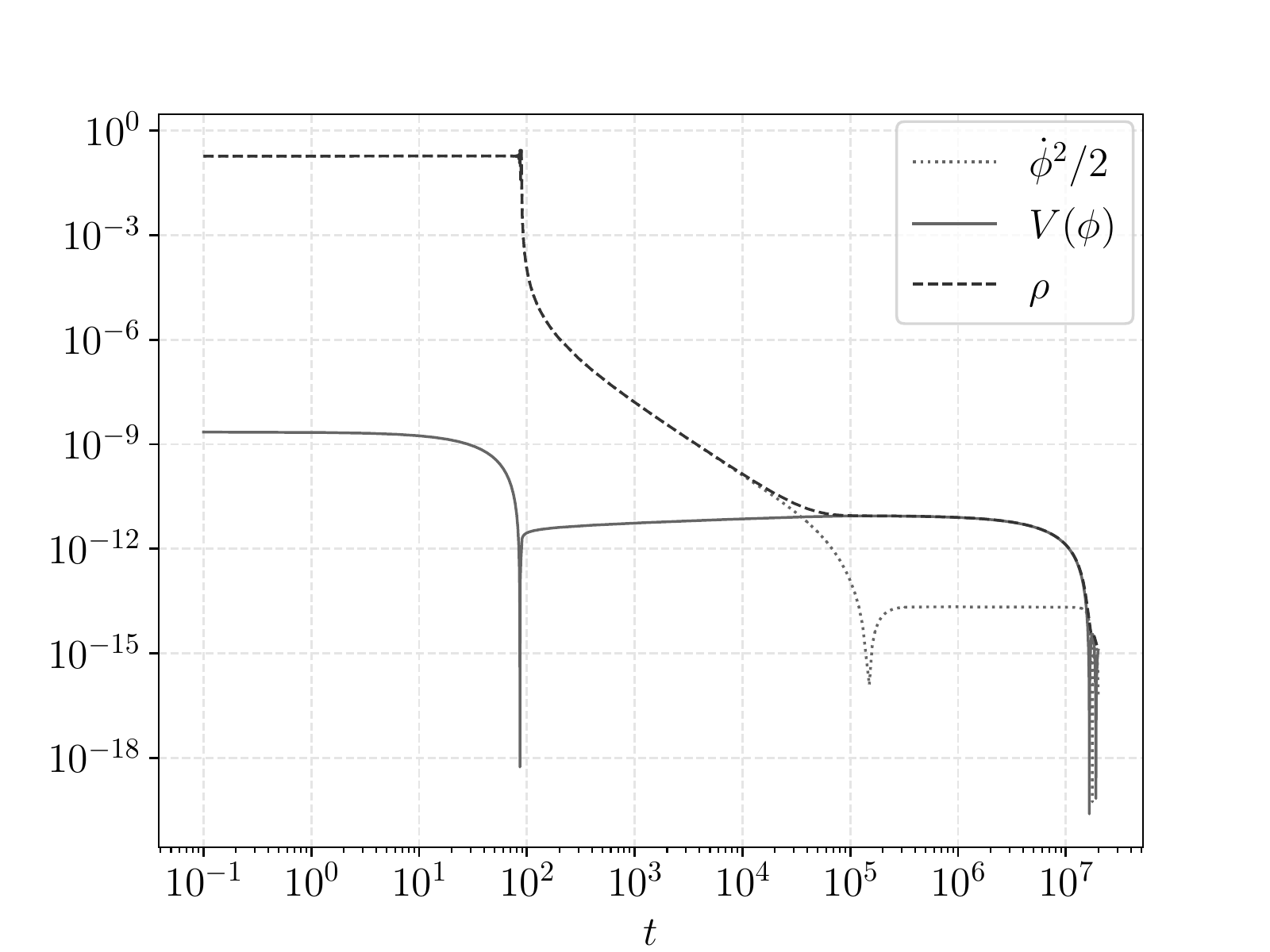}  
}
\caption{This graph shows the kinetic, potential and the total energy density of the scalar field for the effective dynamics of our model. The mass in the potential is $m=1{.}27795\cdot 10^{-6}$. The initial data at $t_0=0$ is $v(t_0)=125.0$, $b(t_0)=1.09\cdot10^2$, $\phi(t_0)=-57.265$ and $\dot\phi(t_0)>0$ with its value determined by solving the Hamiltonian constraint. }
\label{fig:energy}
\end{figure}

In the following, we will focus on kinetically dominated trajectories at times when the universe transitions from the quantum phase to the classical one. We give an example in Fig. \ref{fig:energy}. There, we identify the quantum and classical regimes, where the (slow-roll) inflationary period takes place until the scalar field reaches the bottom of the potential when the reheating period begins. 

\section{Cosmological perturbations}\label{sec:perts}

\subsection{Classical field equations}

In order to supplement our model with inhomogeneous quantum fluctuations in the geometrical and matter sectors, we will follow the standard procedure in primordial cosmology: incorporate inhomogeneous degrees of freedom by means of classical perturbations around FRW geometries, and eventually adopt a quantum representation for them \cite{bran,lan}. We will briefly summarize the main well-known standard steps followed in LQC, which follow several parallelisms with the classical treatment. The reader interested in more concrete details can consult Refs. \cite{drsd,hyb5}. 

Given the classical background metric $h_{ij}^{(0)}$, which is determined by a fiducial metric and the scale factor $a$ (which satisfies Einstein equations), we introduce perturbations around these spacetimes as
\begin{equation}
h_{ij}(t,\vec x) = h_{ij}^{(0)}(t)+\epsilon \delta h_{ij}(t,\vec x)+\ldots,
\end{equation}
where $\epsilon$ is a perturbative (dimensionless) parameter and $\delta h_{ij}$ is the first order perturbation of the metric. For the lapse and shift we adopt a similar expansion 
\begin{equation}
N(t,\vec x) = N^{(0)}(t)+\epsilon \delta N(t,\vec x)+\ldots, \quad N^i(t,\vec x) = \epsilon \delta N^i(t,\vec x)+\ldots,
\end{equation}
with $N^{(0)}$ the zero order homogeneous lapse function, and $\delta N$ and $\delta N^i$ the perturbations of the lapse and shift, respectively. Finally, for the matter sector, the scalar field is decomposed into its homogeneous part and perturbations as follows
\begin{equation}
\Phi(t,\vec x) = \phi(t)+\epsilon \delta \phi(t,\vec x)+\ldots
\end{equation}
In all these expressions the dots represent higher order perturbations that will be neglected in the following. Besides, the perturbations are pure inhomogeneous fields, namely, their zero modes are all equal to zero. Similar considerations apply to the momentum variables conjugate to the previous configurations.

We now introduce these expressions in the Einstein-Hilbert action and truncate it to second order in the perturbations. Within the Hamiltonian framework, the perturbed action is a combination of constraints: one homogeneous scalar constraint (which carries contributions quadratic in the perturbations), one inhomogeneous scalar constraint and three inhomogeneous momentum constraints, all linear in the perturbations \cite{lan,hyb5}. 

Then, following the standard treatment, we carry out a Fourier expansion of these perturbations as
\begin{equation}
\delta F(t,\vec x) =\frac{1}{v_0}\sum_{\vec {\rm k}}\delta F_{\vec {\rm k}}(t)\,e^{i \vec {\rm k}\cdot \vec x},
\end{equation}
where $\vec {\rm k}=(2\pi/l_0)\{ n_1,n_2,n_3\}$ and $n_i\in\mathbb{Z}$ such that $\vec {\rm k}\neq\vec 0$. 


For the perturbations, besides the Fourier expansion, it is convenient to adopt (due to the symmetries of the background spacetime) the standard scalar-vector-tensor decomposition, namely, we decompose all our perturbations (configuration and momenta) in those parts that behave as scalars, vectors and tensors under rotations around the vector $\vec {\rm k}$.

Within this Fourier space, the constraints linear in the perturbations become algebraic equations that are easy to solve. They generate perturbative diffeomorphisms. In order to deal with this set of transformations rigorously both in the classical and the quantum theory, we will follow the usual strategy where one adopts a description in terms of gauge-invariant perturbations (with respect to the linear constraints). In the canonical framework, we achieve this description naturally by means of a canonical transformation that explicitly splits the inhomogeneous sector of the phase space into the pure gauge and gauge-invariant perturbations \cite{lan,hyb5}. For convenience, we will adopt a description in terms of the well-known Mukhanov-Sasaki variables $\nu$, $\mu_{+}$ and $\mu_{\times}$.  

The equations of motion of the Fourier modes of these perturbations are
\begin{align}\label{eq:h-efft2}
\mu_{\vec {\rm k},\epsilon}''+\left[{\rm k}^2-\frac{a''}{a}\right]\mu_{\vec {\rm k},\epsilon}=0,\\\label{eq:h-effs2}
\nu_{\vec {\rm k}}''+\left[{\rm k}^2-\frac{a''}{a}+\mathcal{U}\right]\nu_{\vec {\rm k}}=0,
\end{align}
where the prime denotes derivative with respect to conformal time, ${\rm k}^2=\vec {\rm k}\cdot \vec {\rm k}$, $\epsilon=\{+,\times\}$ indicates the polarizations for the tensor modes and 
\begin{align}\label{eq:Ueff}
\mathcal{U}=a^2\left[\frac{\partial^2 V(\phi)}{\partial\phi^2}+48\pi G V(\phi)+ 6\frac{a'{\phi}'}{ a^{3}\rho}\frac{\partial V(\phi)}{\partial\phi}-\frac{48\pi G}{\rho} V^2(\phi)\right].
\end{align}
One can easily see, modulo the equations of motion of the scalar field as well as the Friedmann and Raychaudhuri equations of the classical background, that\footnote{Although in the classical theory the right and left-hand sides of Eq. \eqref{eq:q-vs-class} are equivalent on-shell, this is not the case if the background equations incorporate quantum corrections. However, in some particular scenarios \cite{hyb-pred,drsd1} these differences do not seem to affect considerably their predictions.}
\begin{equation}\label{eq:q-vs-class}
\frac{a''}{a}-\mathcal{U} = \frac{z''}{z},\quad z= \frac{a\dot\phi}{H}.
\end{equation}

The form of Eqs. \eqref{eq:h-efft2} is not causal. It is equivalent to the ones typically used in the literature of cosmological inflation. However, this particular form allows us to see easily several properties. The first one is that they are well defined everywhere in the phase space of the homogeneous sector (background variables), except at $a=0$ (classical singularity). The wave equations of the scalar perturbations seem to be problematic also at $\rho=0$ (static solution). However, in that situation, the potential and its derivative with respect to $\phi$ also vanish, and therefore the equations of motion of the scalar perturbations should reduce to the ones of a massless field on Minkowski. Except for these particular situations (and assuming that none of these phase space variables blows up at some finite time), the wave equations determine the scalar and tensor modes at any time. Besides, let us also notice that, when the potential $V(\phi)$ and its derivatives with respect to $\phi$ are small or vanish, the equations of motion of the two type of perturbations agree very well or are identical, i.e. $\cal U$ can be neglected or is identically zero, respectively.

\subsection{Fluctuations of the geometry}

Our purpose is to study how quantum inhomogeneous fluctuations of geometry and matter are affected by the very early evolution of the effective emergent universe spacetime. We will assume that a Fock quantization of classical perturbations propagating on these effective geometries will capture the main properties of those fluctuations. We are not deriving the equations of motion of the perturbations from any effective action. Instead, we postulate them, guided by the following consistency conditions:\\
\\
{\it i}) The dynamics of each scalar and tensor mode, codified in smooth, second order and linear ordinary differential equations, should be well defined everywhere,\\
{\it ii}) the perturbations should be chosen such that they admit a unitary evolution in a standard Fock quantization,\\
{\it iii}) at late times, when the effective quantum geometry is well approximated by a classical FRW spacetime, the equations of motion of the modes should agree with the classical ones,\\
{\it iv}) the wave equations of the scalar and tensor perturbations are selected such that they differ in a potential $\cal U$ that vanishes when $m=0$,\\
{\it v}) at very early times, the wave equations of the tensor modes should converge (dynamically) to the ones of a massless scalar field on Minkowski,\\
{\it vi}) provided that all the previous requirements are satisfied, the functions of the homogeneous phase space variables in the equations of motion of the scalar and tensor modes should be as simple as possible.\\

Conditions {\it i}) and {\it iii}) are minimum consistency requirements. In the case of condition {\it i}), although in the classical theory some popular choices for scalar perturbations do not necessarily satisfy it (for instance the comoving curvature perturbation at the onset of inflation), there are consistent choices. Condition {\it ii}) allows us to restrict the study to the Mukhanov-Sasaki variables for scalar and tensor perturbations (see Refs. \cite{fmov,uniq-gen} for a detailed discussion). These variables satisfy condition {\it i}) in a flat FRW classical spacetime -- see Eqs. \eqref{eq:h-efft2} and \eqref{eq:h-effs2}. Condition {\it iv}) is satisfied in the classical theory.  The naive interpretation is the following: tensor perturbations can be understood as small inhomogeneous geometrical anisotropies of a massless scalar field (the shear scalar). There is no reason to think that these small geometrical inhomogeneities should change this character even in the deep quantum regime. Condition {\it v}) is satisfied in the classical theory, in absence of matter content, because in this case, the solution to the Einstein equations is just the flat geometry. Similarly, in the deep quantum regime of our effective geometry, the scale factor converges to a constant value, and therefore to a flat geometry. Finally, condition {\it vi}) prevents the addition of quantum corrections that would be otherwise arbitrary without a fundamental justification. These consistency conditions might not fix the equations of motion completely. However, one concrete example that agrees with them is the one given in Eqs. \eqref{eq:h-efft2} and \eqref{eq:h-effs2}. In the following, we will restrict the study to them.\footnote{Let us notice that in Ref. \cite{barrau} only tensor perturbations where studied in some detail. Actually, the equations of motion of those modes agree with the ones we have chosen here.}

One of the main purposes of this manuscript is to compute the power spectra of cosmological perturbations on this emergent universe background at the end of inflation, starting their evolution at very early times. We will then discuss the standard Fock quantization adopted for them. Let us start with the scalar perturbations. The operators corresponding to each mode are
\begin{equation}
\hat \nu (\eta,\vec x) =\frac{1}{v_0}\sum_{\vec {\rm k}}\hat \nu_{\vec {\rm k}}(\eta)\,e^{i \vec {\rm k}\cdot \vec x}=\frac{1}{v_0}\sum_{\vec {\rm k}}\left(\hat a^{(s)}_{\vec {\rm k}}\nu_{{\rm k}}(\eta)+(\hat a^{(s)}_{\vec {\rm k}})^{\dagger}\nu^*_{{\rm k}}(\eta)\right)\,e^{i \vec {\rm k}\cdot \vec x}.
\end{equation}
They are determined by the creation and annihilation operators $\hat a^{(s)}_{\vec {\rm k}}$ and $(\hat a^{(s)}_{\vec {\rm k}})^{\dagger}$, respectively, fulfilling the commutation relations $[\hat a^{(s)}_{\vec {\rm k}},(\hat a^{(s)}_{\vec {\rm k}'})^{\dagger}]= v_0\hbar\delta_{\vec {\rm k},\vec {\rm k}'}$. The Fourier modes $\nu_{k}(\eta)$ form a complete orthonormal basis of complex solutions to the equation \eqref{eq:h-effs2}, normalized to
\begin{equation}
\nu_{{\rm k}}(\eta)(\nu_{{\rm k}}'(\eta))^*-(\nu_{{\rm k}}(\eta))^*\nu_{{\rm k}}'(\eta)=i.
\end{equation}
As it is well-known, the choice of orthonormal basis is tantamount to the choice of vacuum state of this quantum field theory. Moreover, since this orthonormal basis is completely determined, at a given initial time $\eta_0$, by the initial data $\nu_{{\rm k}}(\eta_0)$ and $\nu_{{\rm k}}'(\eta_0)$, which can be parametrized (up to an irrelevant initial global phase) as 
\begin{equation}
\nu_{{\rm k}}(\eta_0)=\frac{1}{\sqrt{2D_{\rm k}}},\quad \nu_{{\rm k}}'(\eta_0)=\sqrt{\frac{D_{\rm k}}{2}}(C_{\rm {\rm k}}-i),
\end{equation}
with $D_{\rm k}\in\mathbb{R}^+$ and $C_{\rm k}\in\mathbb{R}$, $\forall {\rm k}$, we safely conclude that the information of the initial state is completely encoded in these two parameters. Similar arguments apply to tensor perturbations $\hat\mu_{\times}$ and $\hat\mu_{+}$.

Now, we should notice that these perturbations behave as Klein-Gordon fields with a time-dependent potential. As we saw in Sec. \ref{sec:effective}, to the future of the last bounce, the equations of motion of the perturbations agree very well with those of the standard gauge-invariant perturbations in general relativity. On the other hand, in the asymptotic past (prior to the last bounce), the potential \eqref{eq:Ueff} is the only nonvanishing term. From the numerical analysis of the previous section, it is not difficult to realize that the potential $V(\phi)$ and its derivative with respect to the scalar field will be oscillatory functions with frequency (in proper time) given by $m$, while its second derivative is constant and equal to $m^2$. Therefore, although the geometry reaches a constant scale factor, the equations of the scalar perturbations still see a time-dependent potential. Interestingly, one should expect that the oscillatory nature of the potential in Eq. \eqref{eq:Ueff} will induce a parametric resonance with the frequency of the order of $m$ on these scalar perturbations at very early times. Actually, this resonance will affect the modes $k$ of the order or smaller than $m$ during an infinite period of time and will create instabilities, that will require a description beyond our linear approximation. However, let us notice that the typical scales of the Planckian universes that we will consider here, namely the values of $v_0^{1/3}$, will be several orders of magnitude smaller than the scale given by the mass of the inflaton. Therefore, in all our simulations and without introducing any approximation, the scalar modes with physically relevant wavelengths will behave in the asymptotic past as tensor modes, i.e. as massless test fields on a Minkowski spacetime. From the perspective of cosmological perturbation theory, the Hubble horizon of the spacetime is larger than its physical size, and therefore, perturbations will not be affected by its evolution at very early times. In conclusion, our perturbative approach will remain valid.  Regarding tensor perturbations, they  behave as massless test fields on a Minkowski spacetime in the asymptotic past -- see condition {\it iv)} at the beginning of this section. Therefore, they will never be affected by those parametric resonances, regardless of the value of $v_0$.   

Now,  we must recall that modes with comoving wavenumbers of the order or smaller than $\sqrt{|a''/a|}$ will always be affected during their evolution. For instance, our perturbations evolve from a very early epoch in a spacetime with an oscillatory Hubble parameter. These oscillations, with a frequency of the order of the Planck scale, will considerably affect Planck order comoving scales if the amplitude of the former is sufficiently high. This is actually the case few Planck seconds before the last bounce, when the magnitude of the Hubble parameter reaches its maximum amplitude in the quantum region. Therefore, additional features (parametric resonance) in the power spectra at those scales should be expected. Then, the evolution continues into a classical spacetime dominated by the kinetic energy of the scalar field before inflation begins (we do not consider other situations in this manuscript). Then, once the kinetic energy density decreases sufficiently such that the potential energy dominates, inflation begins. In order to compare our results with the ones provided by observations, we will match our comoving scales with the observational ones by means of a rescaling ${\rm k}(k)=lk$, where $l$ is an appropriate constant that will depend on the number of $e$-folds of expansion during inflation, and therefore on the window of observable modes in the CMB. The Planck mission provides a concrete window of comoving wavenumbers denoted by $k$ and measured in the units $(\rm Mpc)^{-1}$. We will choose the number of $e$-folds such that features of the power spectra that break scale invariance appear around $\ell\simeq 20$, or $k\simeq 1{.}4\cdot 10^{-3} \,(\rm Mpc)^{-1}$.

More precisely, in order to select appropriate initial data, we proceed as follows. For the choice of the mass of the scalar field that we consider here, we compute an effective trajectory specifying initial data at early times, when our spacetime has an approximately constant scale factor. At this time, $t_0=0$, we specify $v(t_0)$, $b(t_0)$ and $\phi(t_0)$, and then $\dot\phi(t_0)>0$ by solving the Hamiltonian constraint. We must recall that the scale factor $a$ and the physical volume $v$ are related at any time by $v=v_0a^3$, where $v_0=v(t_0)$ is the physical volume of the universe in the quantum region, namely $a(t_0)=1$. This implies that $v_0$ enters here as another parameter of the model. Besides, it introduces a cutoff in the possible wavenumbers since there will not be comoving wavelengths larger than $v_0^{1/3}$ in this spacetime.\footnote{A recent discussion of compact flat spatial slices in the context of LQC can be found in Ref. \cite{eli-men}.} Therefore, we will focus on comoving wavenumbers ${\rm k}\geq 2\pi/v_0^{1/3}$. Then, we evolve the system (background and perturbations) and compute the power spectra some $e$-folds after the modes of interest cross the horizon. In order to match the scalar power spectrum with observations, we look for the value of $\phi(t_0)$ (or equivalently the number of $e$-folds of inflation) such that: I) the computed scalar power spectrum agrees with the reference one given in Eq. \eqref{eq:planck}, and II) the first deviations from scale invariance occur at scales $k\simeq 1{.}4\cdot 10^{-3}(\rm Mpc)^{-1}$, or equivalently $\ell\simeq 20$. 

Now, we specify the initial vacuum state of these cosmological perturbations as follows. As we just mentioned above, scalar and tensor perturbations behave at very early times as massless Klein-Gordon fields on a Minkowski spacetime. This particular feature provides a natural solution to the initial value problem for cosmological perturbations in this model. Any exact solutions of the wave equations must converge to a linear combination of complex plane waves $e^{\pm i \omega t+\vec {\rm k} \cdot \vec x}$ at very early times. Therefore, the natural choice of positive frequency solutions in our model is just the one that converges in the past to the Minkowski vacuum of a massless scalar field. For scalar perturbations, it is determined by the (normalized) initial conditions
\begin{equation}
\nu_{\rm k}(\eta_0)=\frac{1}{\sqrt{2 {\rm k}}},\quad \nu_{\rm k}'(\eta_0)=-i\sqrt{\frac{{\rm k}}{2}},
\end{equation}
(or equivalently $D_{\rm k}={\rm k}$ and $C_{\rm k}=0$) where $\eta_0$ is some initial time well in the asymptotic past.  For tensor perturbations, we assume the same initial conditions. 

\subsection{Primordial power spectra}

Given the previous equations of motion and initial data, we have computed the power spectra of scalar and tensor perturbations at the end of inflation, or more precisely, several $e$-folds after the modes cross the horizon. For this purpose, it is convenient to define there the comoving curvature perturbation $\hat{\cal R}_{\vec {\rm k}}=\frac{H}{a\dot\phi}\hat\nu_{\vec {\rm k}}$ and the tensor perturbations $\hat h_{\vec {\rm k},\times}=\mu_{\vec {\rm k},\times}/a$ and $\hat h_{\vec {\rm k},+}=\mu_{\vec {\rm k},+}/a$. Then, the power spectra are obtained by means of the Fourier transforms of the 2-point functions in the real space of each type of perturbation, and they are defined by
\begin{align}
P_{\cal R}(t,{\rm k}) &= \hbar\frac{{\rm k}^3}{2\pi^2}\left(\frac{H(t)}{a(t)\dot\phi(t)}\right)^2|\nu_{\rm k}(t)|^2,\\
P_{h}(t,{\rm k}) &= \hbar\frac{32 G {\rm k}^3}{\pi}\frac{1}{a^2(t)}|\mu_{\rm k}(t)|^2.
\end{align}

\begin{figure}
{\centering     
\includegraphics[width = 0.49\textwidth]{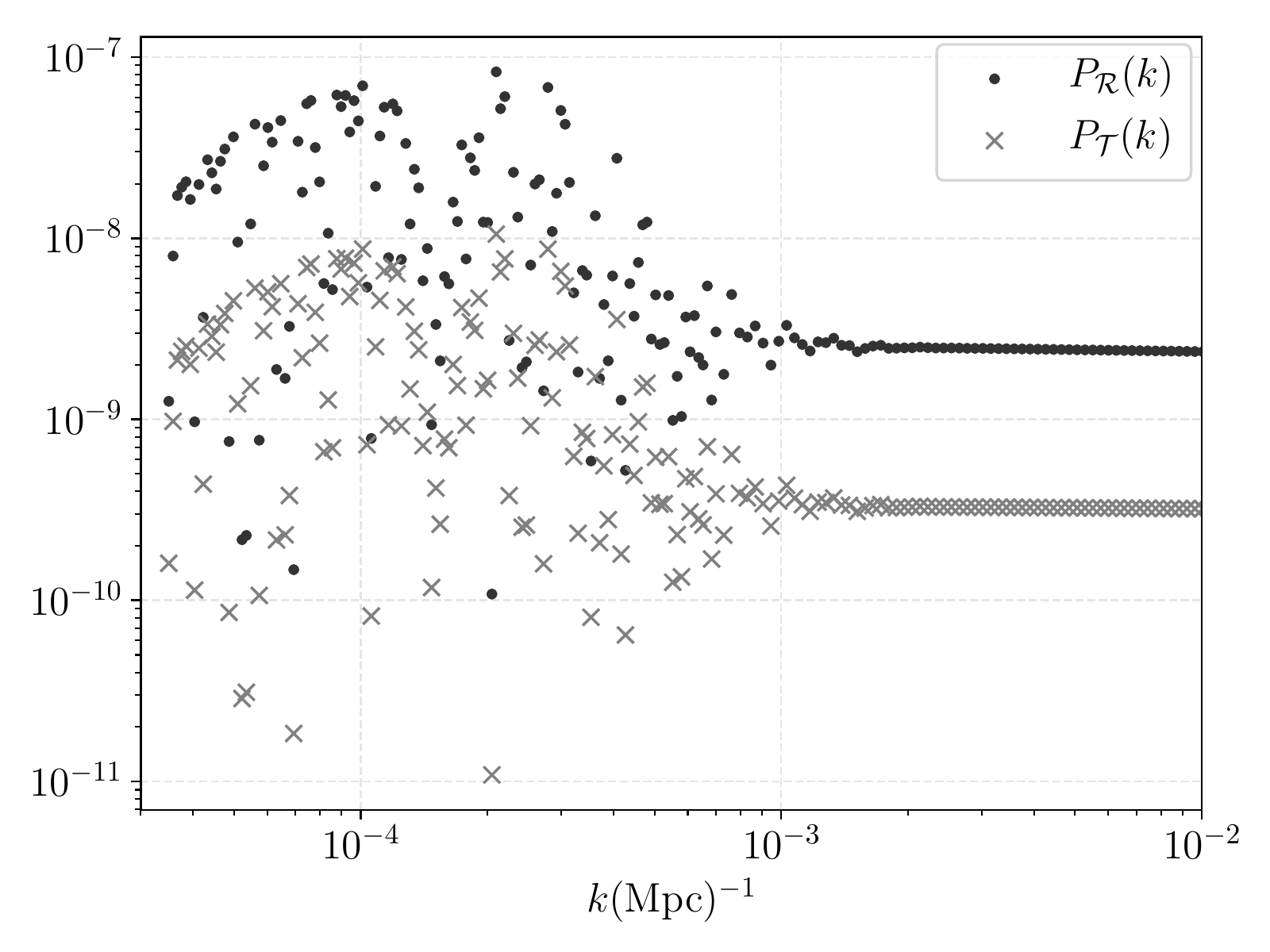} 
\includegraphics[width = 0.49\textwidth]{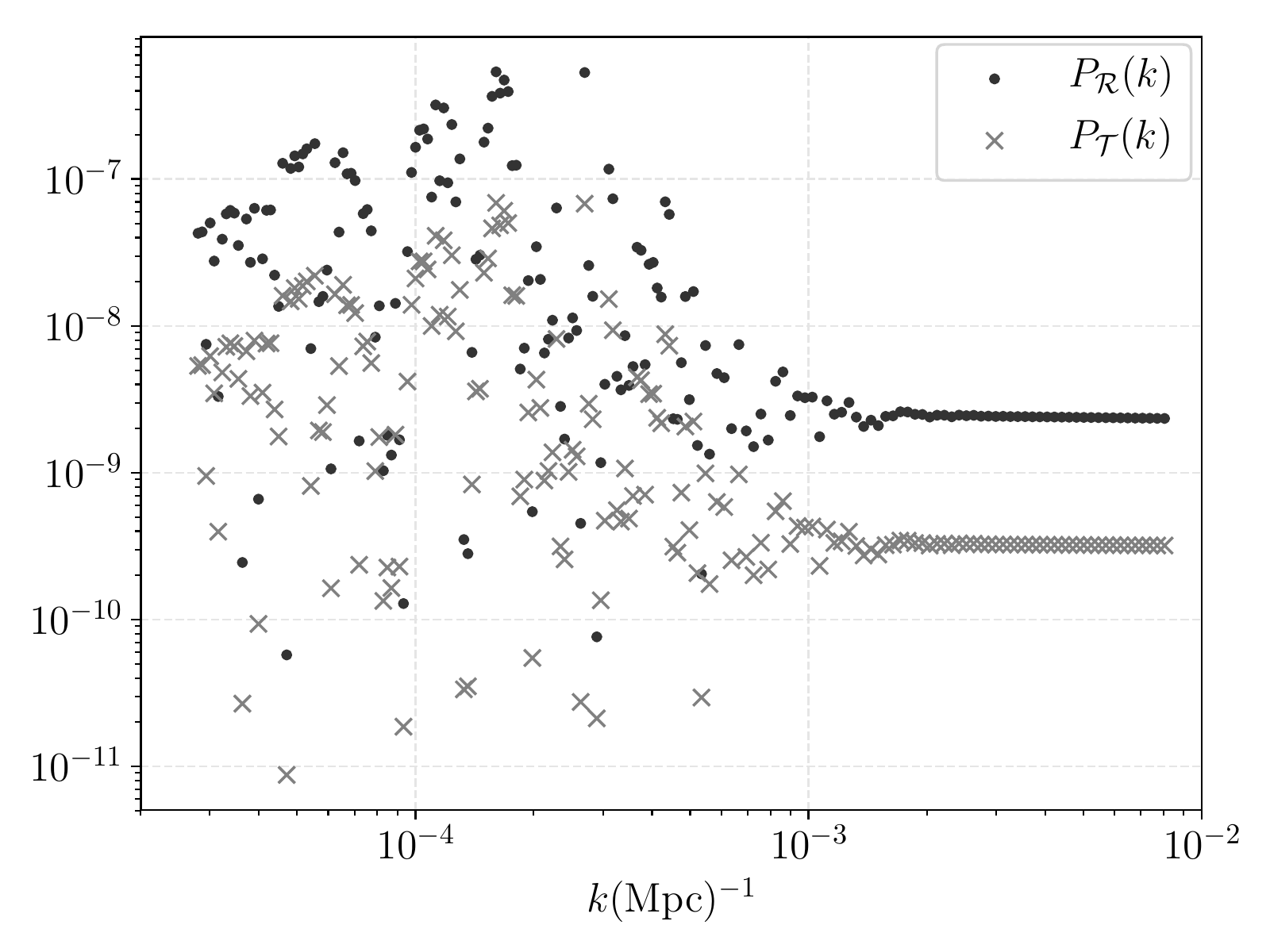}} 
\caption{Scalar and tensor power spectra. The initial data is the following. Left panel: $v(t_0)=125.0$ and $\phi(t_0)=-57{.}265$; right panel: $v(t_0)=343.0$ and $\phi(t_0)=-58{.}25$. In both cases $b(t_0)=1.09\cdot10^2$ and $\dot\phi(t_0)>0$ with its value determined by solving the Hamiltonian constraint. }
\label{fig:PS}
\end{figure}

\begin{figure}
{\centering     
\includegraphics[width = 0.49\textwidth]{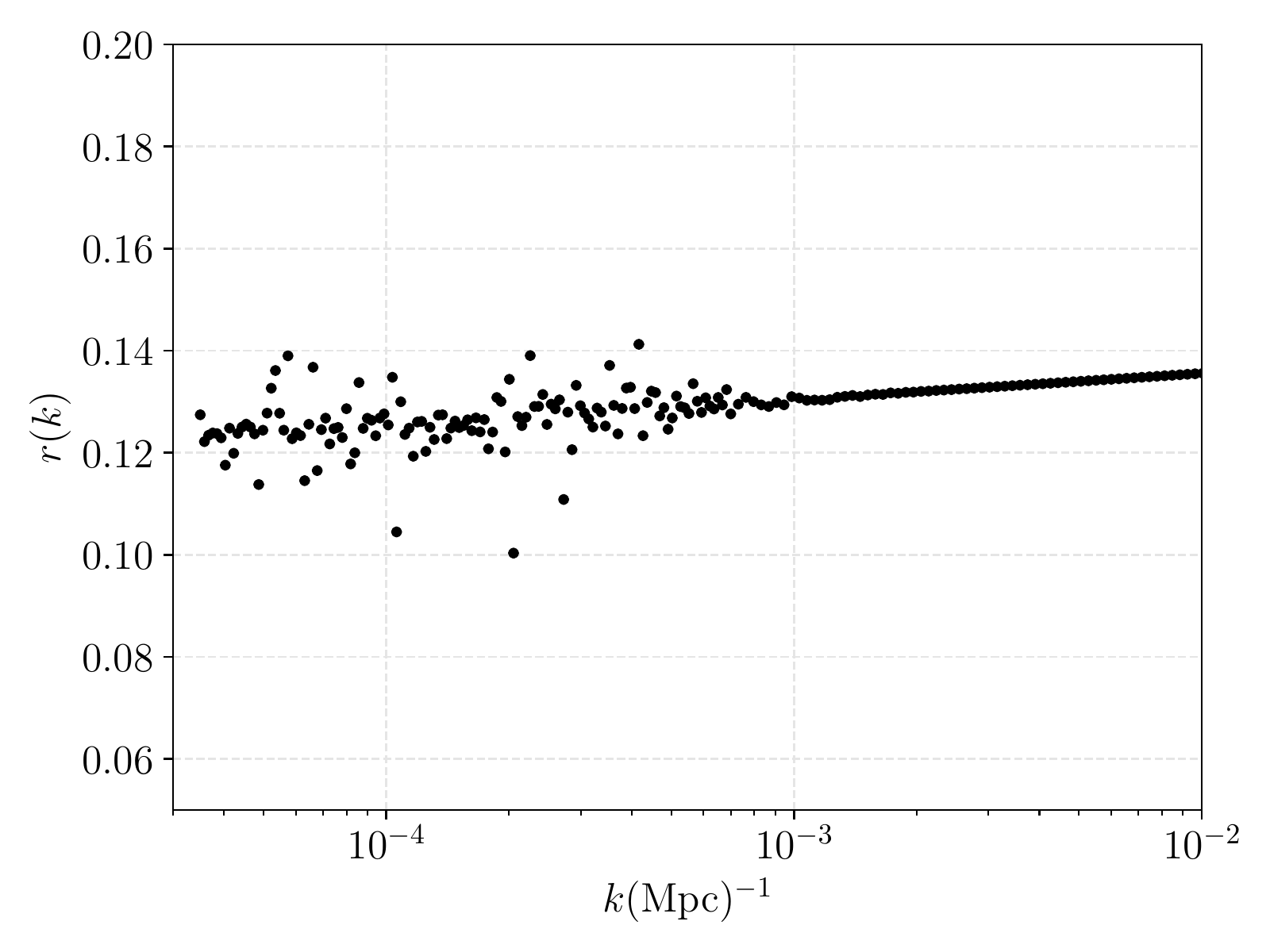} 
\includegraphics[width = 0.49\textwidth]{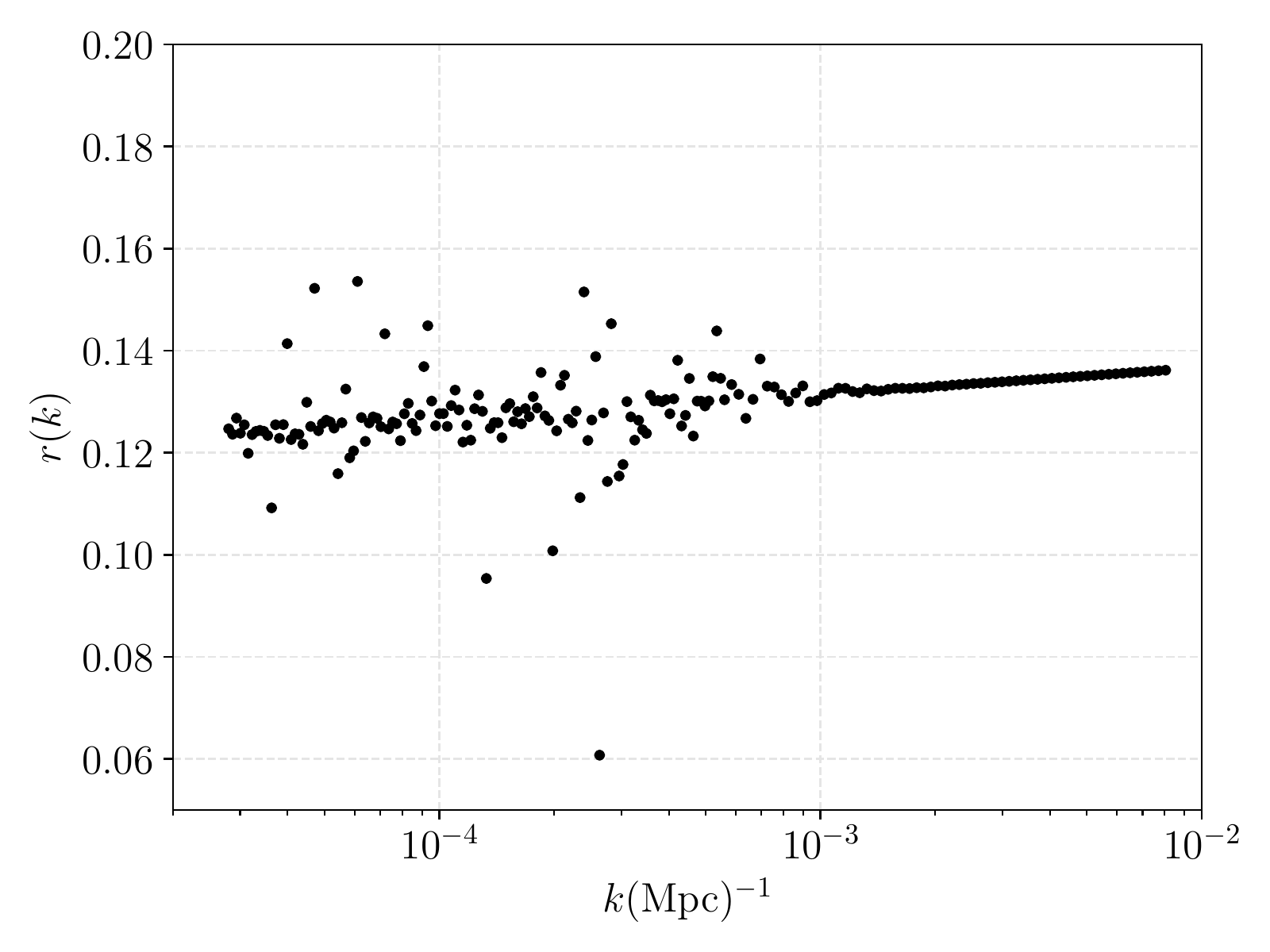}} 
\caption{Tensor-to-scalar ratio. The initial data is chosen as follows. Left panel: $v(t_0)=125.0$ and $\phi(t_0)=-57{.}265$; right panel: $v(t_0)=343.0$ and $\phi(t_0)=-58{.}25$. In both cases $b(t_0)=1.09\cdot10^2$ and $\dot\phi(t_0)>0$ with its value determined by solving the Hamiltonian constraint. }
\label{fig:r}
\end{figure}

Let us now summarize the results of our simulations. Let us recall that we choose the number of $e$-folds such that the scale-invariant region of the power spectrum matches the values provided by the Planck mission, namely, the values given in Eq. \eqref{eq:planck}, and taking into account that the scale invariance is broken around $k\simeq 1{.}4\cdot 10^{-3}({\rm Mpc}^{-1})$, or equivalently, $\ell\simeq 20$ in the multipolar expansion of the temperature anisotropies of the CMB. We then evaluate the scalar and tensor power spectra several $e$-folds after the modes cross the horizon. In Fig. \ref{fig:PS} we show the scalar and tensor power spectra for two different values of the size of the universe in its Planck regime. As we see, both power spectra for very infrared modes (with the smallest wavenumber given by ${\rm k}\simeq 2\pi/l_0$) are enhanced and violate scale invariance. This enhancement is due to a parametric resonance produced at the times where the universe transitions from the deep quantum region to the classical regime. The duration of this transition is more prolonged if the value of $v_0$ increases. On the other hand, for ultraviolet modes, we recover nearly scale-invariant power spectra. The scalar one gives results in agreement with observations. Some qualitative characteristics, like the shape of the enhanced region, are robust with respect to changes in $v_0$. However, this enhancement increases if $v_0$ does. This behavior is also present in the tensor power spectrum. Actually, the relative properties of these perturbations become more clear in Fig. \ref{fig:r}, where we provide the tensor-to-scalar ratio.  For this particular potential, $r(k)$ is approximately constant, even for infrared modes, indicating that the behavior of the scalar and tensor perturbations at early times is very similar. Several models proposed in LQC \cite{hyb-pred2,drsd1} share this feature. However, besides this behavior, we must notice that higher values of $v_0$ introduce larger deviations from a nearly constant tensor-to-scalar ratio. This is a consequence of the fact that the enhancement of infrared modes is very sensitive to the size of the universe in its Planckian regime and because scalar and tensor perturbations do not evolve under exactly the same equations of motion. The magnitude of $r(k)$ for ultraviolet modes is slightly higher than the upper bounds provided by Planck, but this is just a well-known consequence of the concrete quadratic potential selected in the model.

In the light of present and future observations of the CMB, if suppression of power at large scales is confirmed, it seems unlikely that this model will be able to explain it by means of a suppression of power at those scales in the primordial power spectrum. However, this cannot rule out the model since that suppression can be explained by other physical phenomena as well as a sufficiently large number of $e$-folds would hide behind the Hubble radius today the features that break scale invariance in the power spectra. In any case, further research is required in order to understand the model in depth.

\subsection{Comparison with LQC}

We have also computed the power spectrum assuming that the background spacetime is the one provided by the effective equations of motion of LQC. Here, we consider the same mass of the inflaton, we choose the scale factor $a(t_B)=1$, where $t_B$ is the time of the bounce in LQC, and we set $\phi(t_B)$ such that the number of $e$-folds yields a scalar power spectrum in agreement with the estimations provided by Planck mission, given in Eqs. \eqref{eq:planck}, and breaking scale invariance at around $\ell\simeq 20$, or $k=1{.}4\cdot 10^{-3}(\rm Mpc)^{-1}$. We assume the same set of equations for the perturbations as given in Eqs. \eqref{eq:h-efft2} and \eqref{eq:h-effs2}. Their initial state will be the Minkowski vacuum at $10^3$ Planck seconds before the bounce. We also evaluate the scalar and tensor power spectra several $e$-folds after the modes cross the horizon. 

\begin{figure}
{\centering     
\includegraphics[width = 0.49\textwidth]{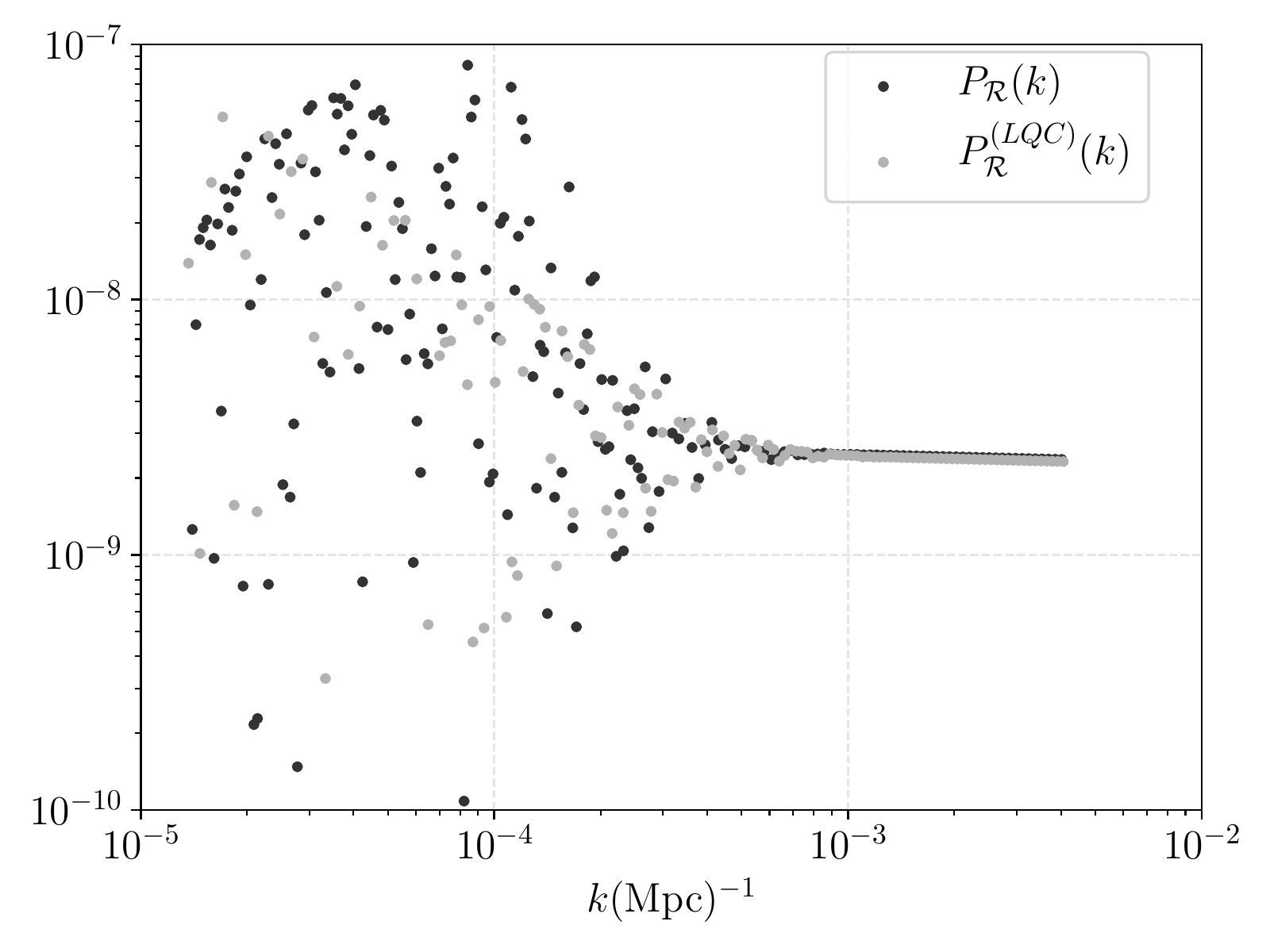} 
\includegraphics[width = 0.49\textwidth]{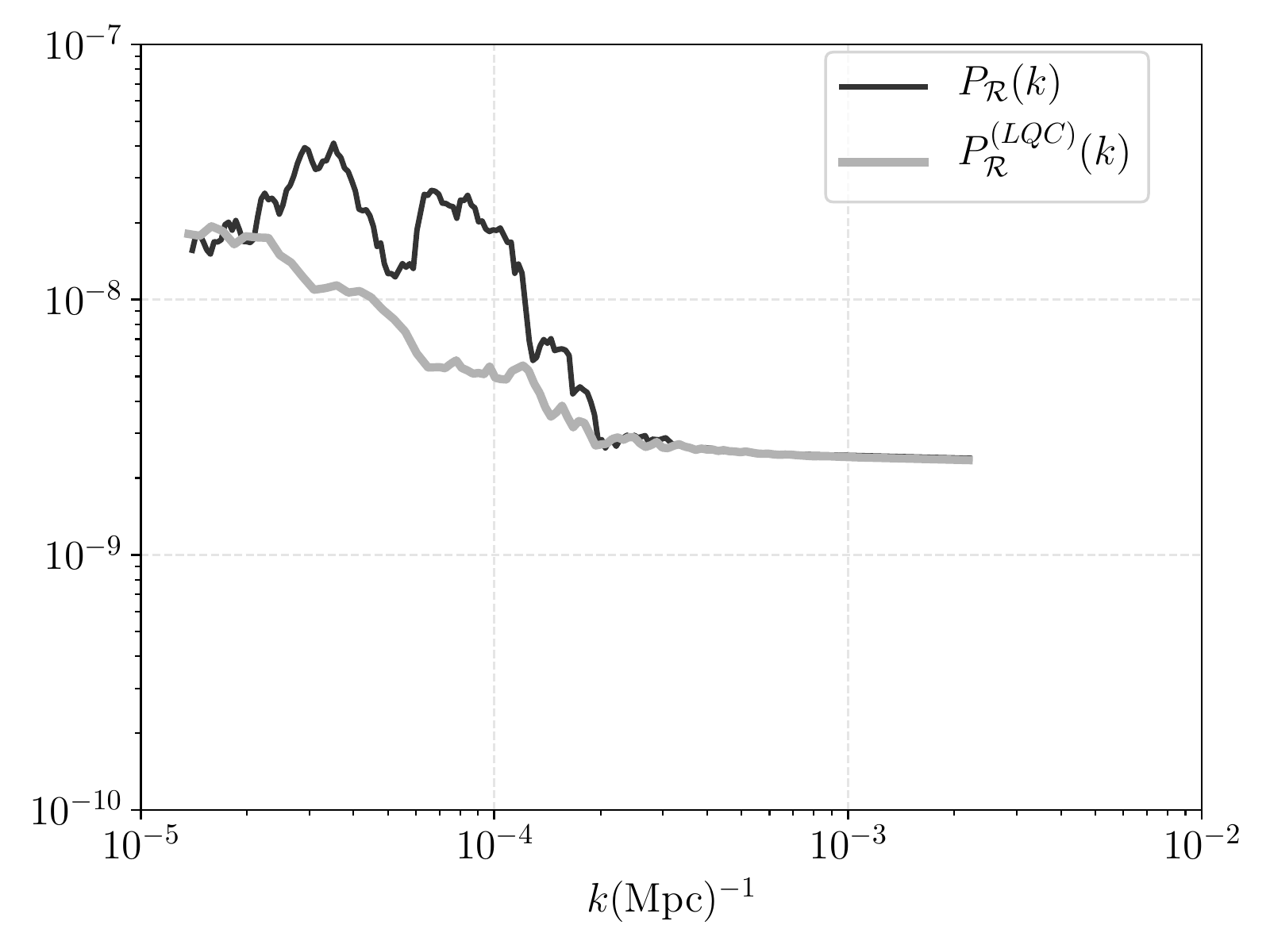}
} 
\caption{Left panel: Scalar power spectrum of our emergent universe model and of LQC. Right panel: Scalar power spectrum of our emergent universe model and of LQC after averaging in a small bin in $k$. The initial data for our background model is: $v(t_0)=125.0$, $\phi(t_0)=-57{.}265$, $b(t_0)=1.09\cdot10^2$ and $\dot\phi(t_0)>0$; the initial data for LQC is $v(t_B)=1.0$, $\phi(t_B)=1{.}077$ and $\dot\phi(t_B)>0$. In both cases the magnitude of $\dot\phi$ is determined by solving the Hamiltonian constraint.}
\label{fig:comp-lqc}
\end{figure}

As we can see in Fig. \ref{fig:comp-lqc}, one of the key differences with respect to loop quantum cosmology is the structure that appears at intermediate scales where the scale invariance of the power spectra is broken, namely, at wavenumbers $k\in[2\cdot 10^{-5},2\cdot 10^{-4}](\rm Mpc)^{-1}$. Here, our model predicts a larger enhancement of power (both for scalar and tensor perturbations) codified in a distinguishable feature naturally explained by the transient parametric resonance at the transition regime, that is absent in LQC. At smaller scales, our model shows agreement with LQC. We have also checked these ratios for other values of $v_0$, obtaining a qualitatively similar behavior. At ultraviolet scales, the two models agree. For tensor perturbations, we observe the same qualitative properties. Hence, we do not show them here. Actually, one could safely conclude this result by noticing that the tensor-to-scalar ratio is constant in average in both LQC and our emergent universe (see Fig. \ref{fig:r}).

\section{Discussion}\label{sec:discuss}

In this manuscript, we have studied some of the most important predictions that can be extracted, from the point of observations, in the context of the emerging universes recently derived within the statistical regularization scheme introduced in quantum reduced loop gravity \cite{rev-qrlg}. Originally, this program derived effective classical Hamiltonians from loop quantum gravity that incorporate nonperturbative quantum corrections. These effective Hamiltonians are determined out of the expectation value of the full Hamiltonian operator on semiclassical coherent states of the full theory. They actually show good agreement with loop quantum cosmology in some regimes. The statistical regularization, a new scheme proposed in Ref. \cite{abcl}, suggests a more natural macrostate constructed out of a superposition of microstates (coherent states) following a suitable probability distribution for the total number of edges of each graph. We have studied several interesting aspects of the resulting effective geometries. In the remote past, the scale factor converges to a constant and, consequently, its time derivatives are zero. Provided suitable initial data, we have seen that the effective equations determine the geometry to the future, showing that after multiple bounces and recollapses, the spacetime geometry can eventually enter the semiclassical regime where classical GR is a very good approximation. These properties are robust for a scalar field subject to a quadratic potential with different values of the mass, including the massless case. We have analyzed the quantum geometry corrections, treating them as an effective fluid. We have seen that the quantum effective energy density and pressure in the remote past coincide with minus the energy density and pressure of the scalar field. Hence, both the Hubble parameter and its time derivative vanish there. We have also seen that for the massive scalar field, once the semiclassical regime is achieved, a trajectory dominated by the kinetic energy density at the beginning of this phase can eventually produce the required number of $e$-folds of inflation. 

Then, we study the evolution of perturbations. We adopt the equations of motion derived from classical theory. Our choice satisfies several consistency conditions suitable for the study of these geometries. They incorporate nonperturbative quantum corrections of the background since, as in the classical theory, they are coupled to the homogeneous sector. In the very early stages of the universe, the modes of these perturbations with physically relevant wavelengths admit natural initial conditions: the ones associated with the vacuum state of a massless scalar field on a Minkowski spacetime. Hence, unlike other cosmological models, our scenario considerably alleviates the ambiguity in the choice of initial vacuum state for perturbations with respect to classical GR or even LQC. The exact positive frequency solutions of our model converge in the past to the ones of a massless scalar field in Minkowski (for the modes that are physically relevant in our study including those that interact strongly with the background geometry). This is one of the most interesting features of this scenario.

Once we provided suitable initial conditions for the system, we evolved both the background and the perturbations until the modes are well inside the Hubble horizon during inflation. There, we computed the power spectra of scalar and tensor perturbations. We considered several sizes of the universe in the remote past, as well as a different number of $e$-folds. We saw that in these cases, the power spectrum is robust for small wavelengths and agrees with observations. However, our model introduces a physical scale that breaks the (near) scale-invariance of the power spectra. It shares this breakdown and the magnitude of its scale with loop quantum cosmology. There, both power spectra are enhanced, but the enhancement is different in these emergent scenarios compared with LQC. It shows a structure (or feature) at some scales similar to the one of a transient parametrized resonance, explained by the fast oscillations of the Hubble parameter and its time derivative in our simulations in the transition between the quantum regime and the semiclassical phase. This structure is imprinted likewise in the scalar and tensor power spectrum. Therefore, the tensor-to-scalar ratio remains constant in average even in those scales where the scale invariance is broken.

In summary, we have shown that the effective Hamiltonian derived in Refs. \cite{abcl,abs} still provides effective geometries within the emergent universe paradigm. It seems robust under different choices of matter content, and more importantly, they admit inflationary solutions. We have also shown that the standard treatment of cosmological perturbation theory is valid here. Moreover, these effective geometries alleviate the initial value problem of the perturbations. Finally, the primordial power spectra obtained by us show additional feature with respect to standard LQC at the largest scales where the scale-invariance is broken.

There are several questions that will deserve attention in the future. From the numerical point of view, our algorithms are efficient and our approach provides physically consistent results. However, it is not clear when the approximation of the discrete summation by a continuous integral in Eq. \eqref{eq:hgr-qrlg} becomes accurate. From a physical point of view, it would be interesting to analyze in more detail the statistical significance of these models with respect to the observational data, following a deeper analysis in the lines of Ref. \cite{barrau} and the traditional Bayesian methods. Moreover, a more fundamental derivation of the effective equations of motion of the perturbations from the full theory would allow us to sharpen our predictions. In addition, the extension to situations where anisotropies are present would provide additional hints about the physics of more general, and therefore, realistic settings, and test the robustness of these emergent universe scenarios in those cases. Finally, implementing the scalar field adopting a polymer representation, and the corresponding effective Hamiltonian within the statistical regularization scheme of QRLG would be another interesting scenario to be explored in the future. 

\section*{Acknowledgments} The Authors are grateful to Ivan Agull\'o, Abhay Ashtekar and Sina Bahrami for useful discussions. This work was supported in part by the NSF grants PHY-1505411, PHY-1552603, NSF-PHY-1603630, the Eberly research funds of Penn State, funds of the Hearne Institute for Theoretical Physics, by Project. No. MINECO FIS2014-54800-C2-2-P from Spain and its continuation Project. No. MINECO FIS2017-86497-C2-2-P, and by Pedeciba (Uruguay).  

\appendix

\section{Saddle-point approximation}\label{app:sp-appr}

Under the saddle-point approximations of Refs. \cite{abcl,abs}, it is possible to find an approximation for the integral in the gravitational part of the Hamiltonian. Concretely, it is possible to arrive at \cite{abs}
\begin{equation}
H_{\rm sp}^{\rm gr}=-\frac{3}{8\pi G \gamma^2}\frac{v\sin^2(\lambda b)}{ \lambda^2} + \frac{\lambda \sin^2(\lambda b)}{48 \pi G \gamma^2} - \frac{\lambda^3 b^2\cos(2\lambda b)}{48\pi G \gamma^2}.
\end{equation}

The equations of motion are easily obtained out of the total Hamiltonian $H_{\rm sp}^{\rm tot}=H_{\rm sp}^{\rm gr}+H^{\rm matt}$, and are given by
\begin{equation}
\dot{v}=-4\pi G\gamma \frac{\partial H_{\rm sp}^{\rm tot}}{\partial b} = \frac{3 v\sin(2\lambda b)}{(2\gamma\lambda)}+\frac{b \lambda^3 \cos(2\lambda b)}{6\gamma} - \frac{\lambda^2 \sin(2\lambda b)}{12 \gamma} - \frac{b^2 \lambda^4 \sin(2\lambda b)} {6 \gamma },
\end{equation}
and
\begin{align}\nonumber
\dot b &= 4\pi G\gamma \frac{\partial H_{\rm sp}^{\rm tot}}{\partial v} = -\frac{3 \sin^2(\lambda b)}{2 \gamma \lambda^2} + 4\pi G\gamma \left(\frac{1}{2}\dot{\phi}^2-V(\phi)\right).
\end{align}

The equations of motion of the matter sector remain unchanged, and are then given by the second order ordinary differential equation \eqref{eq:2nd-diff-phi}.

\section{Numerics}\label{app:numer}

In order to carry out our numerical simulations, we have employed the GNU scientific library. Concretely, for the integration of the ordinary second order differential equations, we need to compute on each time step the integrals given in Eqs. \eqref{eq:I1}, \eqref{eq:I2}, \eqref{eq:I3} and \eqref{eq:I4}. For this purpose, we adopt the adaptive integration procedure QAG. The limits of the numerical integral are selected according to the Gaussian function $e^{-x^2}$ and such that they do not reach values above eight sigmas with respect to it. The QAG algorithm divides the integration interval in subintervals in order to reduce efficiently the overall error. We choose a division in $2000$ intervals. Besides, the absolute and relative errors for this algorithm are set to $1{.}0\cdot 10^{-12}$ and $1{.}0\cdot 10^{-9}$, respectively. Finally, we have checked that our results are robust for the different choices of Gauss-Kronrod rule. Then, the stepping function that we choose for the evolution is an explicit embedded Runge-Kutta Prince-Dormand $(8,9)$ method. We have checked, for several values of the absolute and relative errors, that our results are robust. The ones showed in this manuscript correspond to $h_a=1.0\cdot10^{-14}$ and $h_r=0{.}0$. The initial time step is $t_i=1.0\cdot 10^{-10}$. We have also checked that the magnitude of the (densitized) constraint of the background is compatible with zero, namely $(H^{gr}+H^{matt})\sim 0$ as well as the norm of the modes is not violated, or equivalently 
\begin{equation}
\Im\left[\nu_{k}(\eta)(\nu_{k}'(\eta))^*-(\nu_{k}(\eta))^*\nu_{k}'(\eta)\right]\sim 1,\quad \Re\left[\nu_{k}(\eta)(\nu_{k}'(\eta))^*-(\nu_{k}(\eta))^*\nu_{k}'(\eta)\right]\sim 0,
\end{equation}   
where $\Im[\cdot]$ and $\Re[\cdot]$ mean the imaginary and real parts, respectively.

\end{document}